%% file: main.tex
\pdfoutput=1
\RequirePackage{ifpdf}

\documentclass[cits,11pt,a4paper]{article}
\usepackage{jheppub}

\usepackage{graphicx}

\usepackage{lineno}
\usepackage{amsmath}
\usepackage[toc,page]{appendix}

\input{src/Commands.tex}

\def\bfseries{\fontseries \bfdefault \selectfont \boldmath}

\title{Mitigation of Backgrounds from Cosmogenic $^{137}$Xe in Xenon Gas Experiments using $^{3}$He Neutron Capture}

\input{src/Authors.tex}

\abstract{\Xe{136} is used as the target medium for many experiments searching for \bbnonu. Despite underground operation, cosmic muons that reach the laboratory can produce spallation neutrons causing activation of detector materials. A potential background that is difficult to veto using muon tagging comes in the form of \Xe{137} created by the capture of neutrons on \Xe{136}. This isotope decays via beta decay with a half-life of 3.8~minutes and a \Qb\ of $\sim$4.16~MeV. This work proposes and explores the concept of adding a small percentage of \He{3} to xenon as a means to capture thermal neutrons and reduce the number of activations in the detector volume.  When using this technique we find the contamination from \Xe{137} activation can be reduced to negligible levels in tonne and multi-tonne scale high pressure gas xenon neutrinoless double beta decay experiments running at any depth in an underground laboratory.}


\begin{document}
\maketitle

\section{The NEXT program of high pressure xenon gas TPCs}
\label{sec:NEXT}
The NEXT program has developed the technology of high-pressure xenon gas Time Projection Chambers (TPCs) with electroluminescent amplification (HPXeTPC) for neutrinoless double beta decay searches \cite{Nygren:2009zz, Alvarez:2012haa}. The possibility to achieve sub-1\% FWHM energy resolution and to topologically identify signal-like events was proven in small scale prototypes \cite{Alvarez:2013gxa,Ferrario:2015kta} and has since been tested underground at demonstrator-scale with the NEXT-White (NEW) detector \cite{Monrabal:2018xlr,Ferrario:2019kwg,Renner:2019pfe,Novella:2019cne}. The subsequent stage of the project will deploy 100~kg of \Xe{136} as NEXT-100, currently under construction at Laboratorio Subterr\'aneo de Canfranc (LSC), Spain, with the goal of setting a competitive limit on the \Xe{136} \bbnonu\ half life with the world's lowest background index in xenon.

The future of \bbnonu\ searches involves experiments using one to several metric tonnes of target mass running for tens of years deep underground. A HPXeTPC with a tonne or more of \Xe{136} has great discovery potential given the field's present understanding of neutrino masses. To reach target sensitivities of $10^{28}$ years, improvements over the NEXT-100 background budget \cite{Martin-Albo:2015rhw} will have to be made. Selection of ever purer materials for the construction of detectors enables a considerable reduction in key backgrounds of a radiogenic nature, particularly those from the decays of \Tl{208} and \Bi{214}. As radiogenic backgrounds become sub-dominant, other sources of background become relevant, such as those of cosmogenic origin.  These backgrounds cannot be reduced simply by selection of purer detector materials and must be mitigated by other means.

\begin{figure}[t]
\centering
 \includegraphics[width=0.48\linewidth]{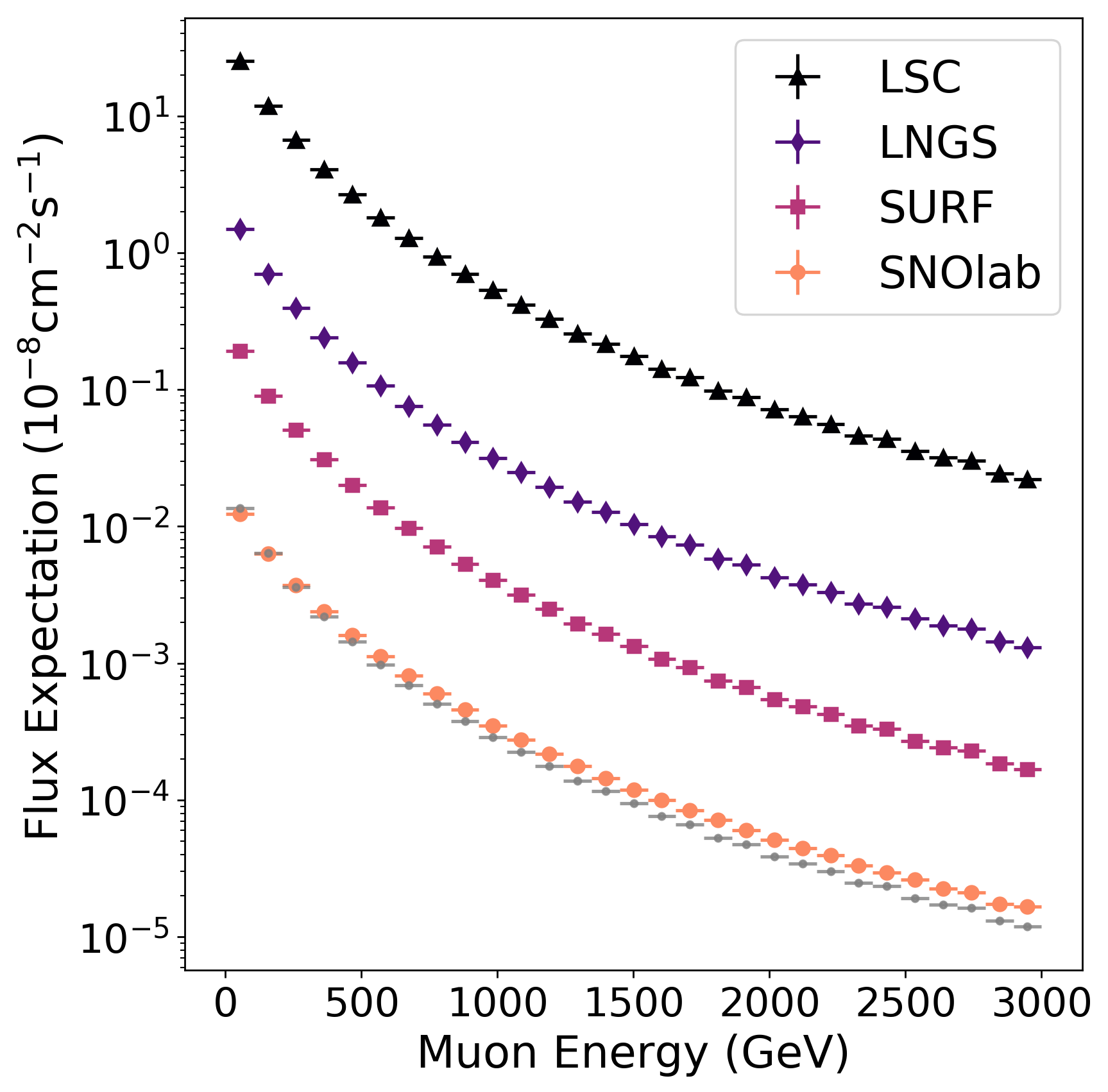}
\includegraphics[width=0.49\linewidth]{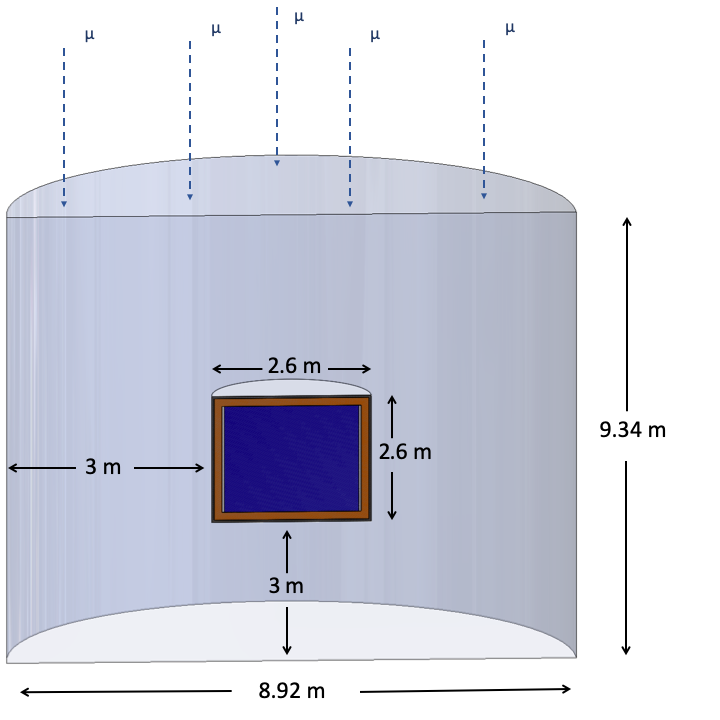}
\caption{\emph{Left}: Expected cosmic muon energy spectrum in four different labs being considered for ton-scale detectors. Grey dots represent scaling LNGS muon distribution with SNOlab's muon flux while the orange are using the MUSIC muon simulation code for SNOlab. \emph{Right}: Geometry used in simulations.}
\label{fig:SimGeom}
\end{figure}

The location for the first module of the tonne-scale NEXT program, called NEXT-HD, has not yet been determined, but various underground labs worldwide are under consideration including SNOLab, LNGS, LSC, and SURF. The deeper the laboratory, the lower the muon flux, as shown in Fig.~\ref{fig:SimGeom}, \emph{left}. A lower muon flux implies lower contamination of cosmogenic backgrounds, which has prompted most experimental neutrinoless double beta decay programs to favor the deepest available sites, with more than 5 km.w.e (kilometers water equivalent) of overburden. 

Since the NEXT detectors operate using xenon gas, it is feasible to mix certain additives into the volume to improve detector properties. Any additive must meet a set of minimal criteria for the experiment to succeed: it must not attach ionization electrons during their drift or interfere with the electroluminescence process, must not absorb scintillation light, and must not negatively affect energy resolution to a substantial degree. The additive must also be in gas phase at room temperature, be compatible with hot and cold getters in the purification system, and it must be possible to circulate it through the gas system. In addition, all considered additives should ideally be non-toxic.

The collaboration is investigating several possible additives which could improve the topological reconstruction when compared to operation with pure xenon, one of which is helium \cite{XePa,Felkai:2017oeq,Henriques:2018tam,McDonald:2019fhy,Fernandes:2019zuz}. The predominant isotope of helium is $^4$He, which, if added in quantities between 10\% and 15\% has a substantial positive impact on transverse diffusion.  The sub-dominant isotope of helium, $^3$He, is present in natural helium at the 2$\times 10^{-4}$ level.   Unlike $^4$He, $^3$He has an extremely high capture cross section for neutrons.  In this paper we consider the positive impacts of adding a small quantity of $^3$He to enriched xenon to dramatically reduce contamination from cosmogenic backgrounds in tonne or multi-tonne scale underground high pressure xenon gas detectors.

One can also consider the use of the technique presented here in liquid xenon experiments.  With a boiling point of 3~K, however, \He{3} will tend to concentrate in the vapor of the ullage in a liquid xenon detector, rather than remaining in the liquid phase.  A minority amount of helium will remain in the liquid, as implied by Henry's law, though the Henry coefficients for helium in xenon are not presently known.  The LUX collaboration has shown that helium can be loaded into liquid xenon at the level of .003 - .009\% by mass \cite{Lippincott:2017yst}, but this level of doping is insufficient to affect a significant reduction in \Xe{137} contamination, based on the studies presented in this work.

\section{Cosmogenic neutron backgrounds}
\label{sec:Motivation}
As radiogenic background sources are reduced and target masses increase, cosmogenic backgrounds become more apparent. The most pernicious of these backgrounds derive from neutrons.  This is because neutrons are very penetrating, and can capture on nuclei to create long-lived beta or gamma emitters, which can decay with a signal in the energy region of interest for double beta decay.   While prompt backgrounds from nuclear cascades post-capture can be effectively vetoed using muon taggers of various types, longer lived isotopes pose a greater threat. In the case of \Xe{136} experiments, the only long lived isotope likely to be produced in laboratory conditions with a decay that can mimic the \bbnonu\ signal is \Xe{137}. The beta-decay of this isotope with \Qb\ of 4.162~MeV produces electrons with a continuum of energies that includes the \Xe{136} Q-value \Qbb~=~2.458~MeV, and can constitute background to the search if not effectively filtered. 

The decay of \Xe{137} has been identified as a significant contributor to the background expectations of several xenon-based double beta decay experiments.  For example, 20 to 30 percent of EXO-200's background was from \Xe{137} \cite{Anton:2019wmi,EXO200::2015wtc}, prompting future liquid xenon TPCs to go very deep underground to escape it. Even at a depth of 6~km.w.e. at SNOLab, the proposed nEXO concept projects a non-trivial background from \Xe{137} \cite{Albert:2017hjq}. KamLAND-Zen has also found a 7\% dead-time from spallation products, the largest source being \Xe{137} with ($3.9\pm 2.0) \times 10^{-3}$ tonne$^{-1}$day$^{-1}$ produced \cite{Gando_2016}.  Finally, it is notable that \Xe{137} activation provides a slow but non-zero source of non-double-beta-decay related barium production through the chain \Xe{137}$\rightarrow^{137}$Cs$\rightarrow^{137}$Ba, a potentially relevant consideration for barium tagging~\cite{jones2016single,mcdonald2018demonstration,byrnes2019barium,Thapa:2019zjk,Rivilla:2019vzd,Woodruff:2019hte}.  In the NEXT detectors these may be largely rejected by time-coincidence cuts with energy deposits of interest as the $Q_{\beta}$ for $^{137}$Cs$\rightarrow^{137}$Ba is lower than the $Q_{\beta\beta}$ of interest.

\begin{figure}[t]
\centering
\includegraphics[width=0.48\linewidth]{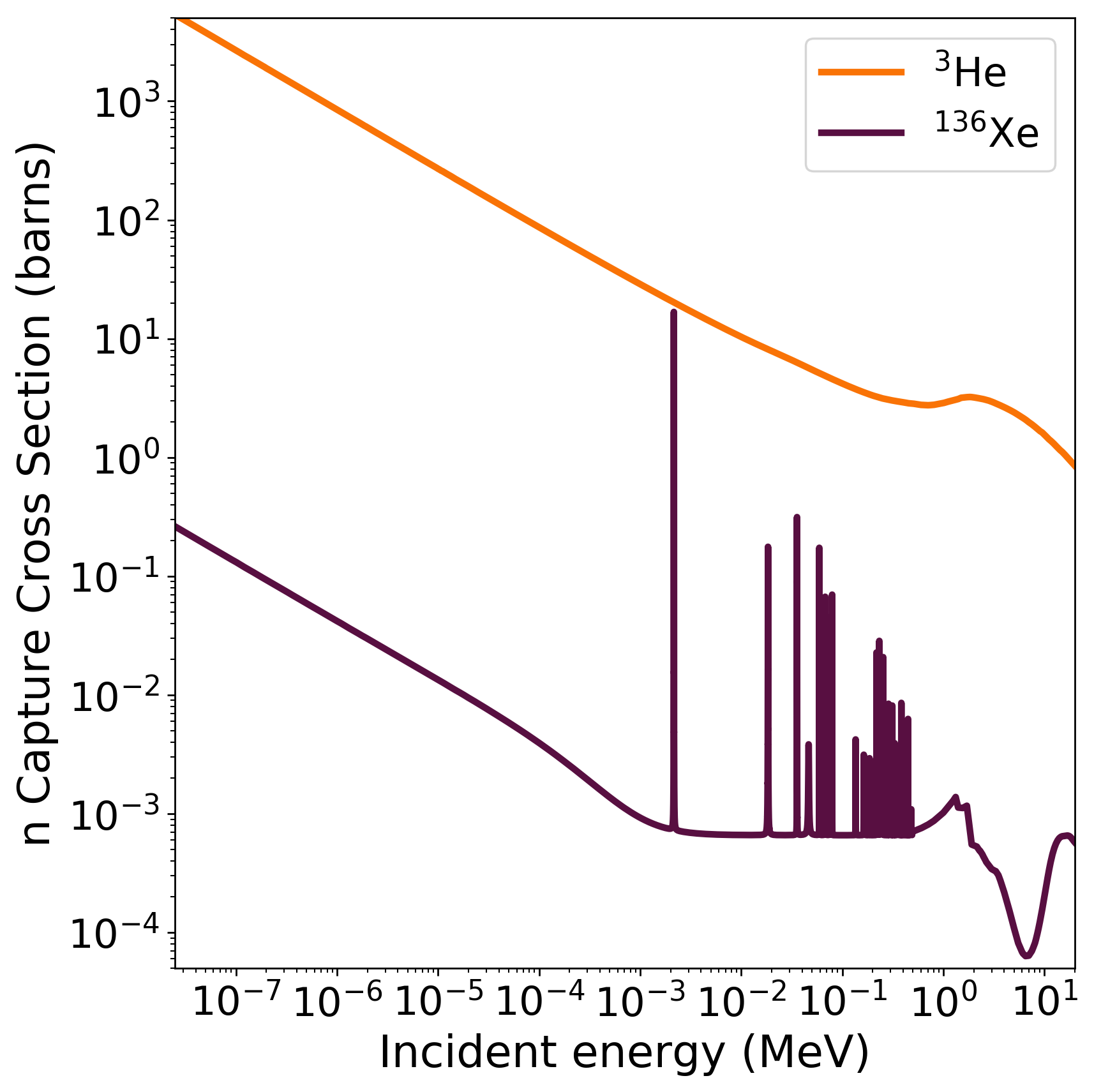}
\includegraphics[width=0.48\linewidth]{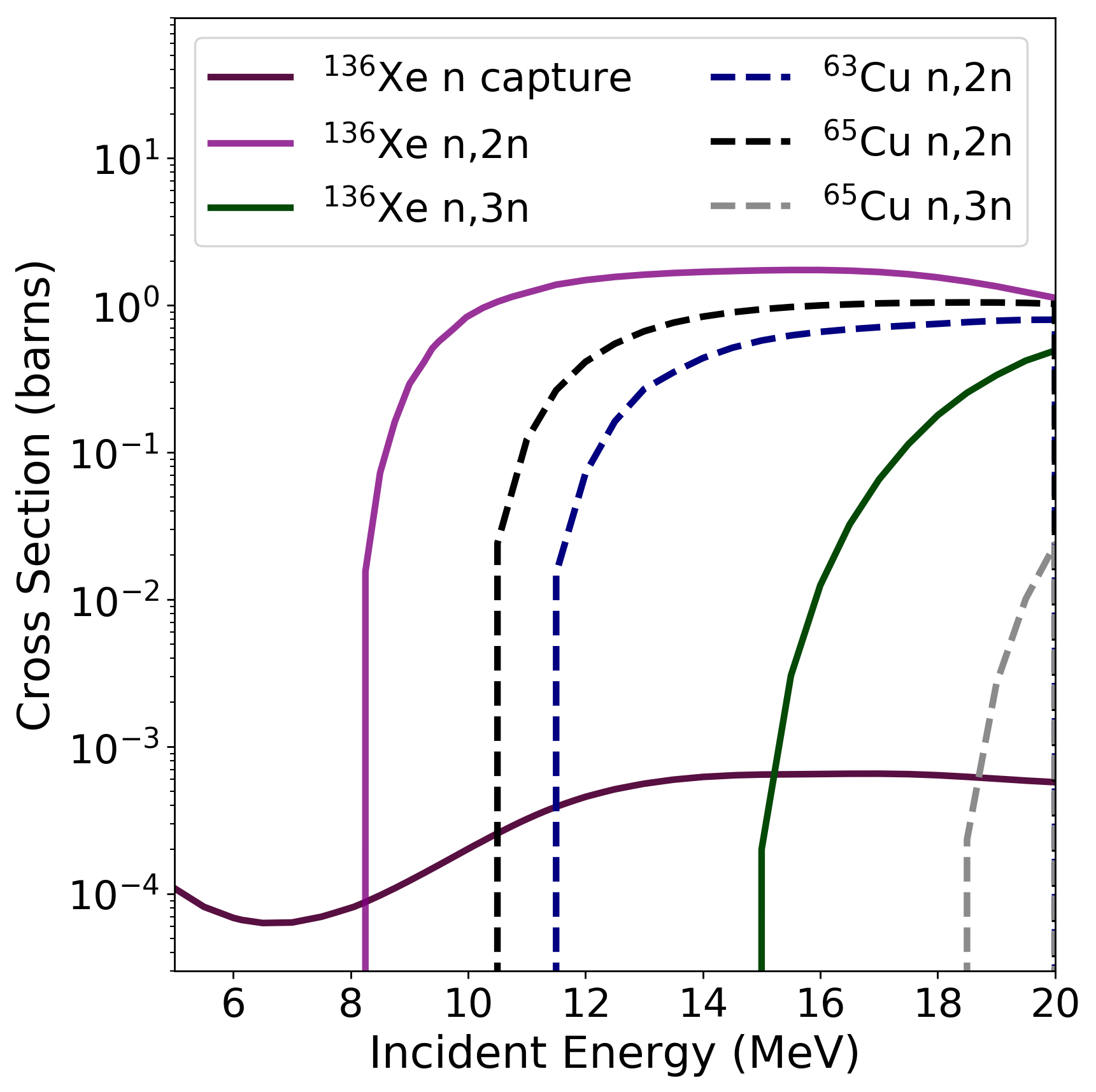}
\caption{Left: Cross sections for neutron capture on \Xe{136} and \He{3}. Right: Cross sections for neutron capture and inelastic scattering on \Xe{136} and the most abundant copper isotopes.  All cross sections are drawn from the ENDF database~\cite{chadwick2011endf}, which mirror those used in {\tt GEANT4}~\cite{Brown:2018jhj}}
\label{fig:crosssections}
\end{figure}

In this work we explore the impact of addition of a small fraction of \He{3} to pure xenon to mitigate  \Xe{137} production and reduce cosmogenic backgrounds for neutrinoless double beta decay. \He{3} has a neutron capture cross-section that is four orders of magnitude greater than that of \Xe{136}, as shown in  Fig.~\ref{fig:crosssections}{\em-left}. The capture process produces hydrogen and tritium and an energy of 764~keV:
\begin{equation}
    ^{3}\mathrm{He} + n \rightarrow\ ^{1}\mathrm{H} +\ ^{3}\mathrm{H}.
    \label{eq:he3capt}
\end{equation}

This process is commonly used in $^3$He-based neutron detectors.     The tritium later beta-decays to \He{3} with \Qb\ of $\sim$18.6~keV with a half life of twelve years. Because their energies are all far below  \Qbb, none of the products of neutron capture on \He{3} present potential background to the \bbnonu\ search.  Backgrounds from tritium contamination at high rates, either as a product of neutron captures or due to contamination of the raw gas (which is typically manufactured through tritium decay) could interfere with detector calibrations that use $^{83m}\mathrm{Kr}$ decay X-rays \cite{Martinez-Lema:2018ibw}.  However, it has been demonstrated that tritium can be effectively removed by getters~\cite{SnoTritium} and so purification both before filling and during detector operations are expected to mitigate this effect.  Moreover, interference with krypton calibration would require pile up of various  tritium decays and is, as such, likely to be a negligible contribution to the high statistics runs used for calibration. By absorbing a large quantity of thermal neutrons without introducing high energy backgrounds, the presence of \He{3} in the active volume is expected to significantly reduce the abundance of neutron captures on xenon.  By effectively mitigating the background from \Xe{137} production, such an admixture could allow high pressure xenon gas detectors to operate at shallower depths, and also provide a tool for monitoring the neutron flux to better understand contributions from other neutron-induced signatures. 

\section{Simulations of \Xe{137} activation}
\label{sec:app}
A Monte Carlo study was carried out using the NEXT simulation framework to investigate the impact of \He{3} doping on cosmogenic background contamination. For this study we consider an HPXeTPC with active volume diameter of 2.6~m and length of 2.6~m at 15~bar giving a total \Xe{136} target mass of $\sim$1109~kg for xenon enriched to 90.3\% in \Xe{136}. The active volume is surrounded by a 1~cm thick plastic cylinder that represents the fieldcage and by 12~cm of copper on all sides. This closely mirrors the proposed geometry of the tonne-scale NEXT-HD, though using a simplified simulation volume.  NEXT-HD will be submerged in an instrumented water tank as a means to mitigate interactions of rock neutrons and to tag cosmic muons. To this end we simulated a cylindrical water tank surrounding the TPC with diameter 9.34~m and a height of 8.92~m to give 3~m of water on all sides of the pressure vessel (see Fig.~\ref{fig:SimGeom} \emph{right}).

The effect of the addition of \He{3} on \Xe{137} activation was investigated in two ways: First, high statistics Monte Carlo sets with neutrons at low energies within the active volume were used to directly study activation (described in section \ref{subSec:intNeut}). Two neutron injection energies were investigated: a) 10~eV neutrons, which quickly thermalize (``thermal''); b) 10~MeV neutrons which may experience harder scatters or inelastic processes before thermalizing (''fast'').  Second, muons with energies between 1~GeV and 3~TeV, representing $\sim$99.5\% of the expected energy range in most underground labs (see Fig. \ref{fig:SimGeom}-\emph{left} for flux expectations), were launched from above the water tank in order to study the expectations from this source both with and without \He{3} additive (section \ref{subSec:Muons}). The simulations were performed using {\tt GEANT4}~\cite{Agostinelli:2002hh} version {\tt 10.5.p01.} In addition, further cross checks were performed using {\tt GEANT4} version {\tt 10.6.p01} as well as {\tt FLUKA} version {\tt 2011.2x-8}  and found to be consistent. The cross checks are summarised in section \ref{sec:Checks}.

The most important cross sections for this work are those relating to thermal neutron capture. The capture cross sections in this version of {\tt GEANT4} are drawn from the {\tt ENDF/B-VII.1} database~\cite{chadwick2011endf}. The ENDF cross sections were derived originally from Ref.~\cite{mughabghab2006atlas} and were recently tested experimentally, with the measured thermal neutron capture cross section on \Xe{136} validated at the 1$\sigma$ (approx. 20\%) level~\cite{Albert:2016vmb}.  Scattering and capture of neutrons on \He{3} has been studied extensively (for example, Refs~\cite{batchelor1955helium,manokhin1988brond,mughabghab2006atlas,gibbons1959total,haesner1983measurement,antolkovic1967study,seagrave1960elastic,sayres1961interaction,als1964slow}) and the ENDF database recommends a sub-percent uncertainty on the provided 3He(n,p)t cross section below 1~keV, growing to 5\% at 50~keV.  Cross sections for neutron production through inelastic scattering are calculated within {\tt GEANT4} using the Bertini~\cite{Bertini:1963zzc,Barashenkov:1972id,Bertini:1970zs} intra-nuclear cascade model, which has been validated against data for inelastic neutron scattering in terms of both angular and energy distribution on a variety of nuclei~\cite{heikkinen2003bertini}, with agreement at the factor-few level~\cite{wright2015geant4}. The central result of this work, the expected improvement in \He{3} / \Xe{136} mixtures over pure \Xe{136} in terms of \Xe{137} production, depends primarily on the ratio of neutron capture cross sections of \He{3} and \Xe{136}. Based on the above considerations, this is expected to be accurate at the 20\% level.  The absolute \Xe{137} yield, however, depend on the details of neutron production in complex showers, and this carries a far larger uncertainty.  This uncertainty provides a further motivation for the use of \He{3} in underground experiments using \Xe{136}, to monitor the total thermal neutron yield, itself proportional to the total rate of \Xe{137} activation in the detector.

\section{Results}
\label{sec:res}

\subsection{\Xe{137} Production from internal neutrons}
\label{subSec:intNeut}

The potential of \He{3} as a neutron-moderating additive in NEXT was first studied using neutrons simulated in the internal volumes of the detector. 10\ensuremath{^6}\ neutrons were simulated with energies \textless 10~eV to check thermal neutron captures, and 10 MeV for fast neutrons, generated over 4~$\pi$ solid angle, starting in the field cage structure with xenon-helium gas mixtures at 15 bar and 300 kelvin. The helium percentage-by-mass ranged between 0 to 5$\%$ and the number of \Xe{137} created were counted for each run. A control simulation set was generated using a mixture of the same enriched xenon with \He{4}. This helium isotope does not capture neutrons and, as such, its admixture is expected to have no effect on the number of \Xe{137} produced. This data set is used to validate that it is indeed the neutron-capturing properties of \He{3} that lead to any observed changes in \Xe{137} background, rather than dilution or the neutron-moderating impact associated with additional light target nuclei.

\begin{figure}[t]
\centering
\includegraphics[width=0.48\linewidth]{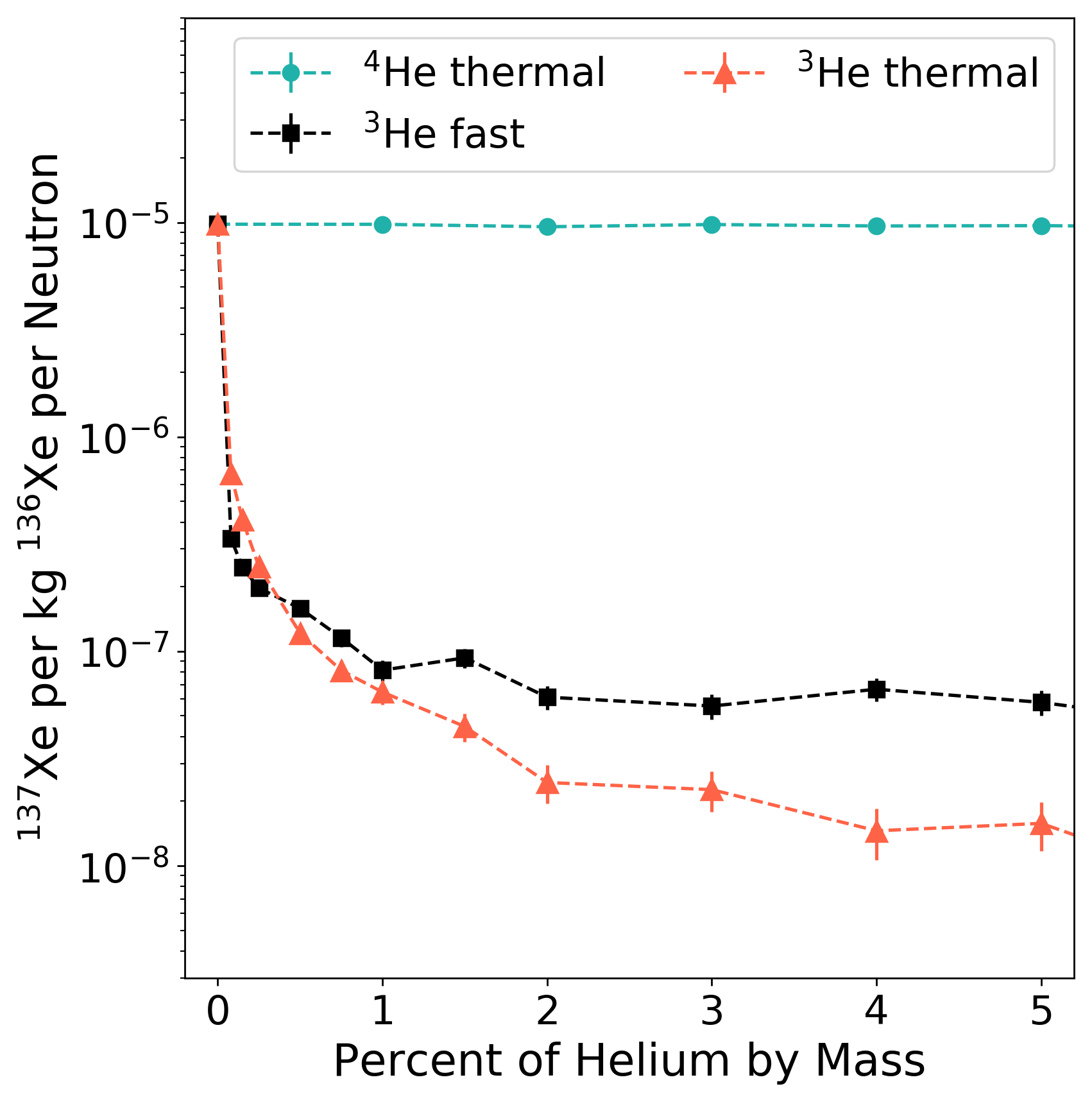}
 \includegraphics[width=0.48\linewidth]{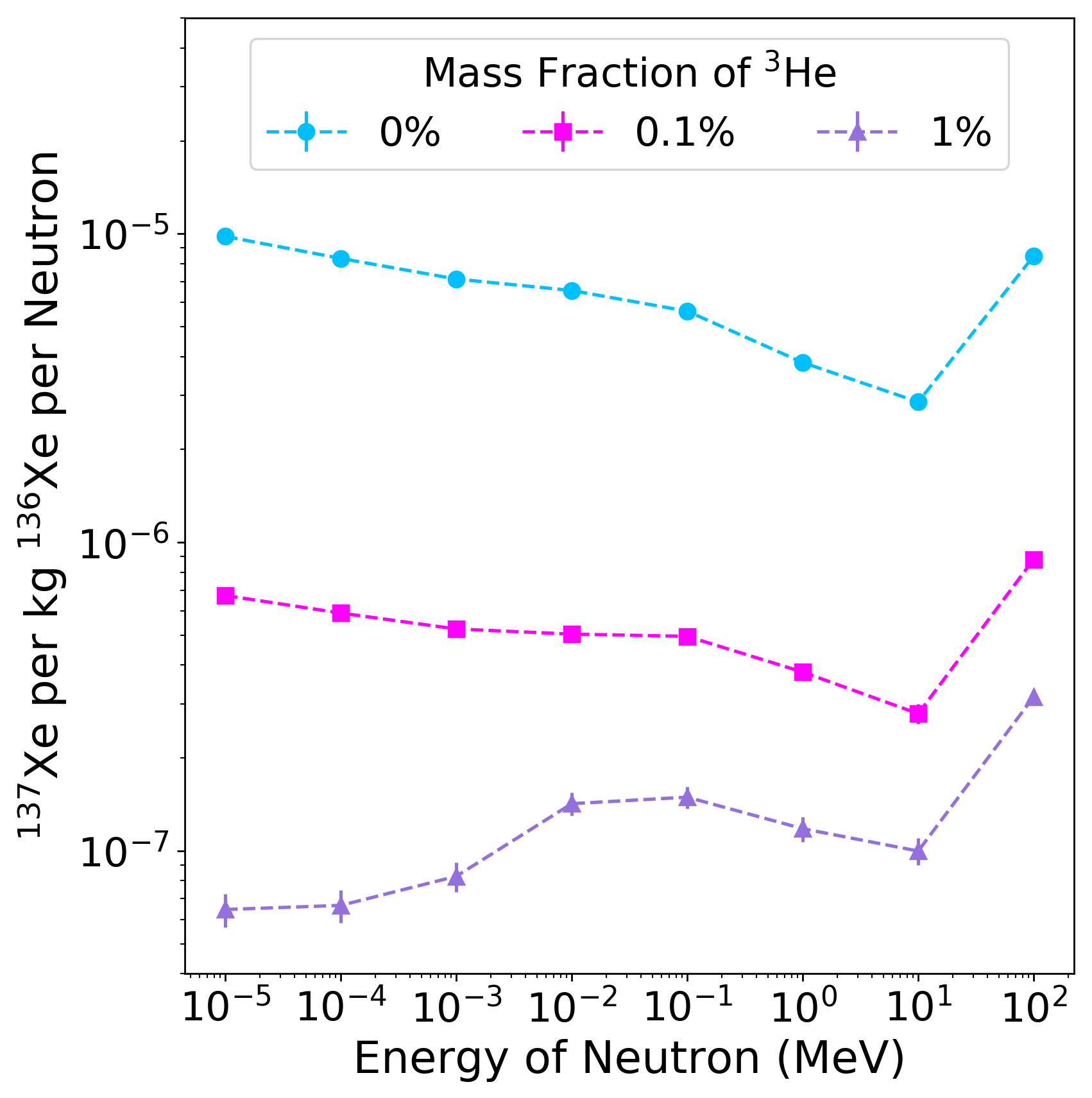}
\caption{Left: Number of \Xe{137} nuclei created by neutron capture with varying amounts of \He{4}  and \He{3}. All thermal neutrons started at 0-10 eV and all fast neutrons started at 10 MeV. Right: Comparison of Helium plus enriched xenon and pure enriched xenon with varying neutron energies.}
\label{fig:percofheliums}
\end{figure}

The number of activations is normalized to the total target mass, i.e, the number of kilograms of \Xe{136} in the active volume. With the largest helium fractions, dilution alone has some small impact on the \Xe{137} rate, which is not the effect we intend to study here.  Using the mass of \Xe{136} as the denominator avoids this issue.  The exact normalization used is

\begin{equation}
    N^{\prime}_{137} = \displaystyle\frac{N_{137}}{E_{136}\cdot P_{Xe}\cdot m_{a}},
    \label{eq:massNorm}
\end{equation}
where $N_{137}$ is the number of \Xe{137} produced in the simulations, $E_{136}$ is the level of enrichment in the 136 isotope, $P_{Xe}$ is the proportion of the gas mixture taken up by xenon, and $m_{a}$ is the mass in the active volume of the detector.

Figure~\ref{fig:percofheliums}-\emph{left} shows the results for mixtures with \He{4} and \He{3}. No statistically relevant change in the normalized number of activations for any proportion of \He{4} is found. This is not the case when we consider an addition of \He{3} to the gas. Figure \ref{fig:percofheliums}-\emph{left} clearly demonstrates the power of the \He{3} to absorb neutrons, and to remove contamination from \Xe{137}. Even at a fraction of a percent concentration, a clear reduction in \Xe{137} activation is seen.  By 0.5\% of \He{3} a reduction of two orders of magnitude is predicted from both thermal and fast neutrons. 

Considering both a 0.1 and 1\% \He{3} addition and varying the initial neutron energy, it can be seen that the number of activations continues to be significantly reduced across all energy bins. Figure \ref{fig:percofheliums}-\emph{right} shows the dependency over several orders of magnitude in neutron energy. Such a reduction even at the 0.1\% level appears sufficient to drive the \Xe{136} background in a ton-scale experiment to negligible levels, even at modest detector depths. We return to this point quantitatively in Sec.~\ref{subSec:Muons}.

There are notable features in the energy-dependencies of Fig.~\ref{fig:percofheliums}-\emph{right}.  The origins of these spectral effects were investigated by detailed examination of the Monte Carlo simulation predictions.  The steady fall in activation as a function of energy between 10~eV and 1~MeV in pure xenon corresponds to the increasing probability that a neutron will leave the active volume without thermalizing as the energy increases.  The neutron capture cross sections are also falling in this region, though the capture is predominantly effective for thermalized neutrons due to the very large number of scatters each neutron can undergo with its surroundings once thermal.  The sharp increase in \Xe{137} production at around 10~MeV observed in all three curves corresponds to the sharp up-tick in rates of multi-neutron production processes at these energies. Both (n,2n) and (n,3n) cross sections on xenon and copper become large in this region, as can be seen in Fig.~\ref{fig:crosssections}, right, reproduced from the ENDF database.  Above 10 MeV, therefore, each primary neutron can be the parent of many more secondary neutrons, leading to enhanced production of \Xe{137}  per injected parent neutron.  The ``bump'' in rate of capture at intermediate energies in the 1\% \He{3} / \Xe{136} mix system is attributed to capture of fast neutrons by resonances in the \Xe{136} neutron capture cross section, shown in Fig.~\ref{fig:crosssections}, left.  In pure xenon and in the 0.1\% \He{3} / \Xe{136} system, where the overwhelming majority of neutrons producing \Xe{137}  are thermal, fast neutron captures are a negligible fraction of the population and no shape effect from these resonances is visible.  In the 1\% \He{3} / \Xe{136} system, with thermal neutrons effectively mitigated by \He{3}, fast neutron capture becomes a more substantial contribution.  This transition explains the initially rapid drop breaking to a slower fall of Fig.~\ref{fig:percofheliums}, left.  The addition of a small quantity of \He{3} quickly absorbs the majority of thermalized neutrons that have been slowed by repeated elastic collisions with \Xe{136}, for both initial injection energies.  At higher concentrations the capture of fast neutrons becomes increasingly relevant, affecting the level at which the capture rate plateaus for high concentrations of \He{3}.

\begin{figure}[t]
\centering
\includegraphics[width=0.49\linewidth]{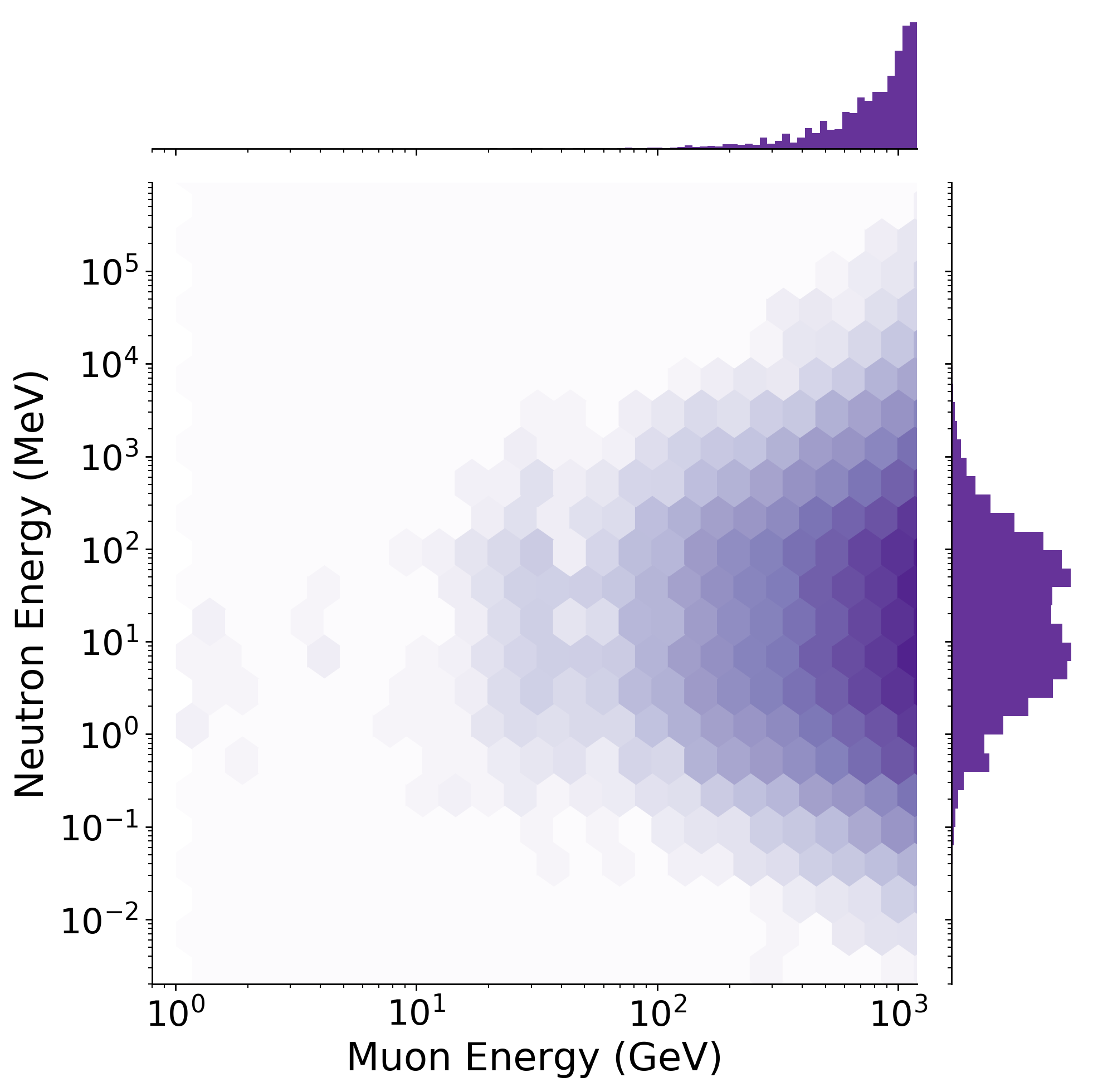}
\includegraphics[width=0.49\linewidth]{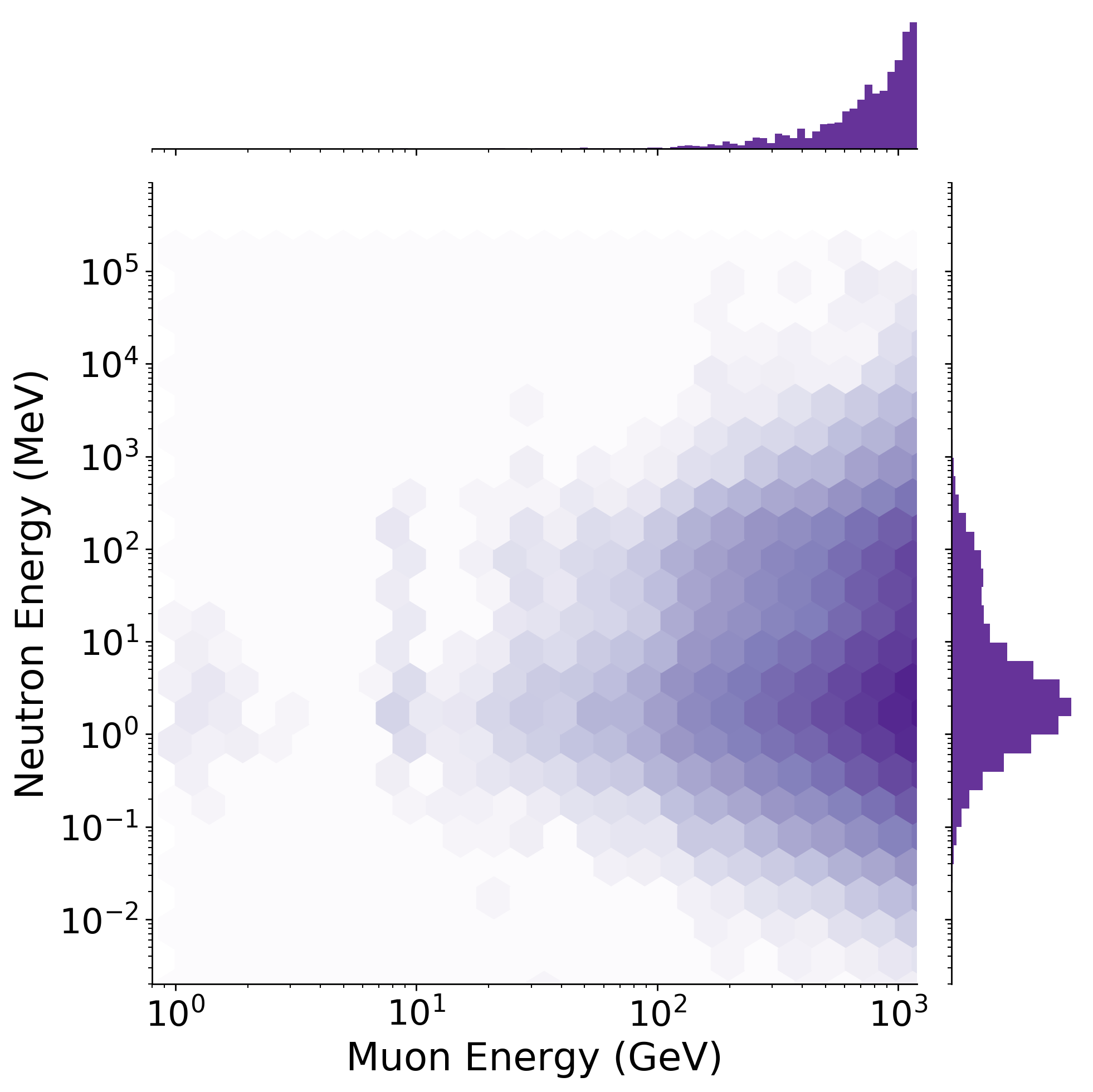}
\caption{\emph{Left}: Correlation plot for neutron and muon energies for neutrons created within the water tank; \emph{Right}: Correlation plot for neutrons created in detector materials.}
\label{fig:NuandMuenergies}
\end{figure}

\subsection{\Xe{137} Production from muons}
\label{subSec:Muons}

The source of neutrons most likely to reach the active volume of the detector comes in the form of spallation products created by muon-material interactions that thermalize in detector materials before being captured by the \Xe{136}.  The showers created by muons are complicated systems, involving cascades of particles and a multitude of inelastic processes.  Shown in Fig. \ref{fig:SimGeom}-\emph{left} is the expected cosmic muon energy spectrum in Laboratori Nazionali del Gran Sasso (LNGS) as calculated using the MUSIC muon simulation code \cite{KUDRYAVTSEV2009339}.  To estimate the flux at different laboratories we have used the spectral shape for LNGS multiplied by the absolute normalization for different labs.  While there is an expected hardening of the muon flux for deeper labs, our comparison of the  SNOLab flux predicted by simply renormalizing LNGS flux with the measured flux from \cite{Aharmim_2009} shows that the effect is negligible within the precision of the present study.

To study the expected impact of the muon-material interactions, a high statistics simulation of muons uniformly distributed between 1~GeV and 3~TeV was performed. The neutron spectrum produced by muons interacting in the detector materials have energies in the range 0.01~MeV~--~100~GeV (see Fig.~\ref{fig:NuandMuenergies}). The distribution of neutron energies depends weakly on the muon energy, but the number of neutrons produced depends on it strongly. Fig.~\ref{fig:NuandMuenergies} \emph{left} gives the energy distribution of neutrons produced within the water tank, and Fig.~\ref{fig:NuandMuenergies} \emph{right} gives the neutrons produced within the detector materials and xenon, but not the water tank.  Additional contributions to the high energy peak in the water tank neutron energy distribution arise from capture processes including $\mu^-+p\rightarrow \nu + n$ and  $\pi^-+p\rightarrow \gamma + n$.  A sub-leading contribution from neutron, pion, and proton inelastic scattering populates the high energy peak in the detector volume.

\begin{figure}[t]
\centering
\includegraphics[width=0.30\linewidth]{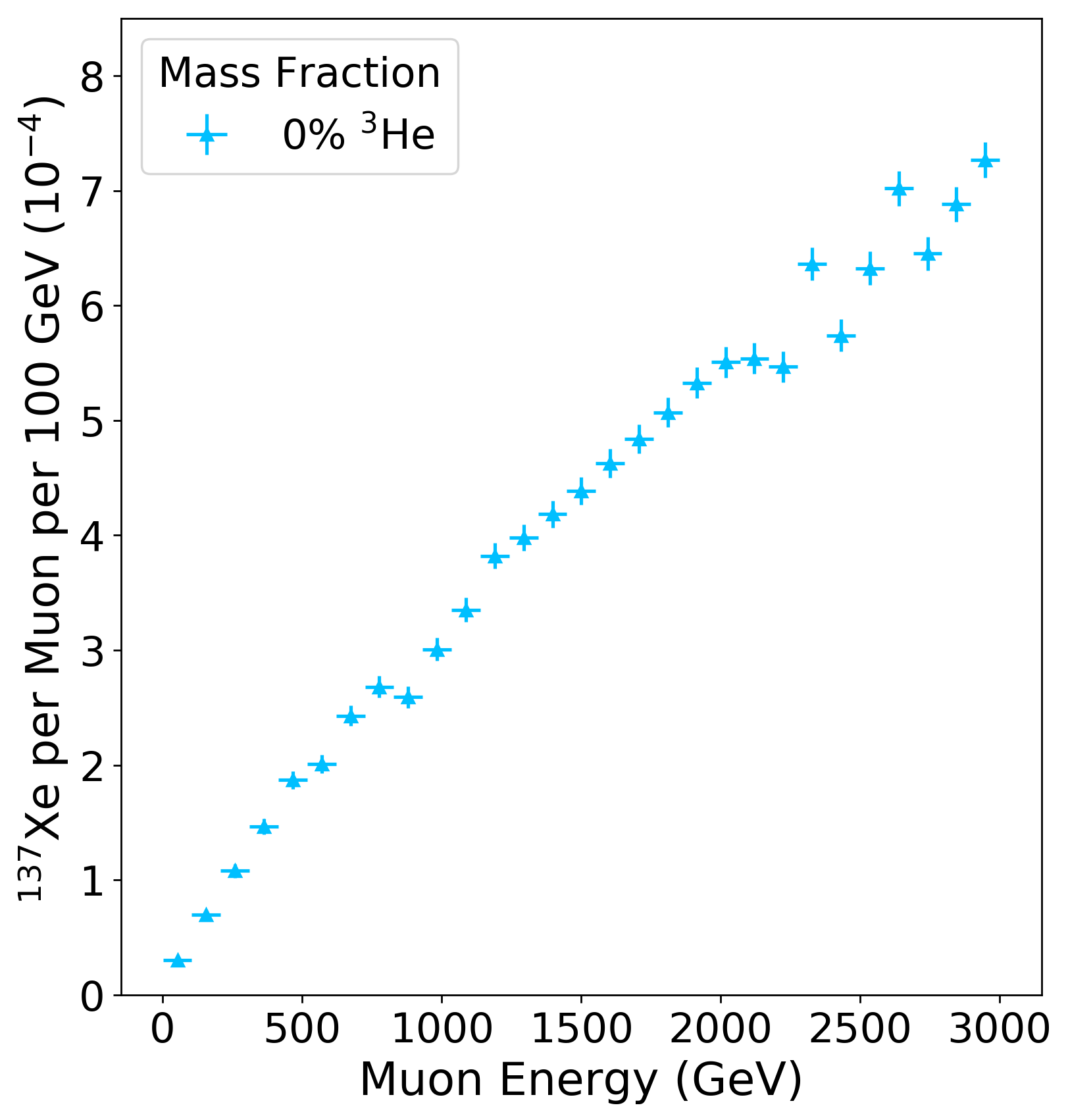}
\includegraphics[width=0.31\linewidth]{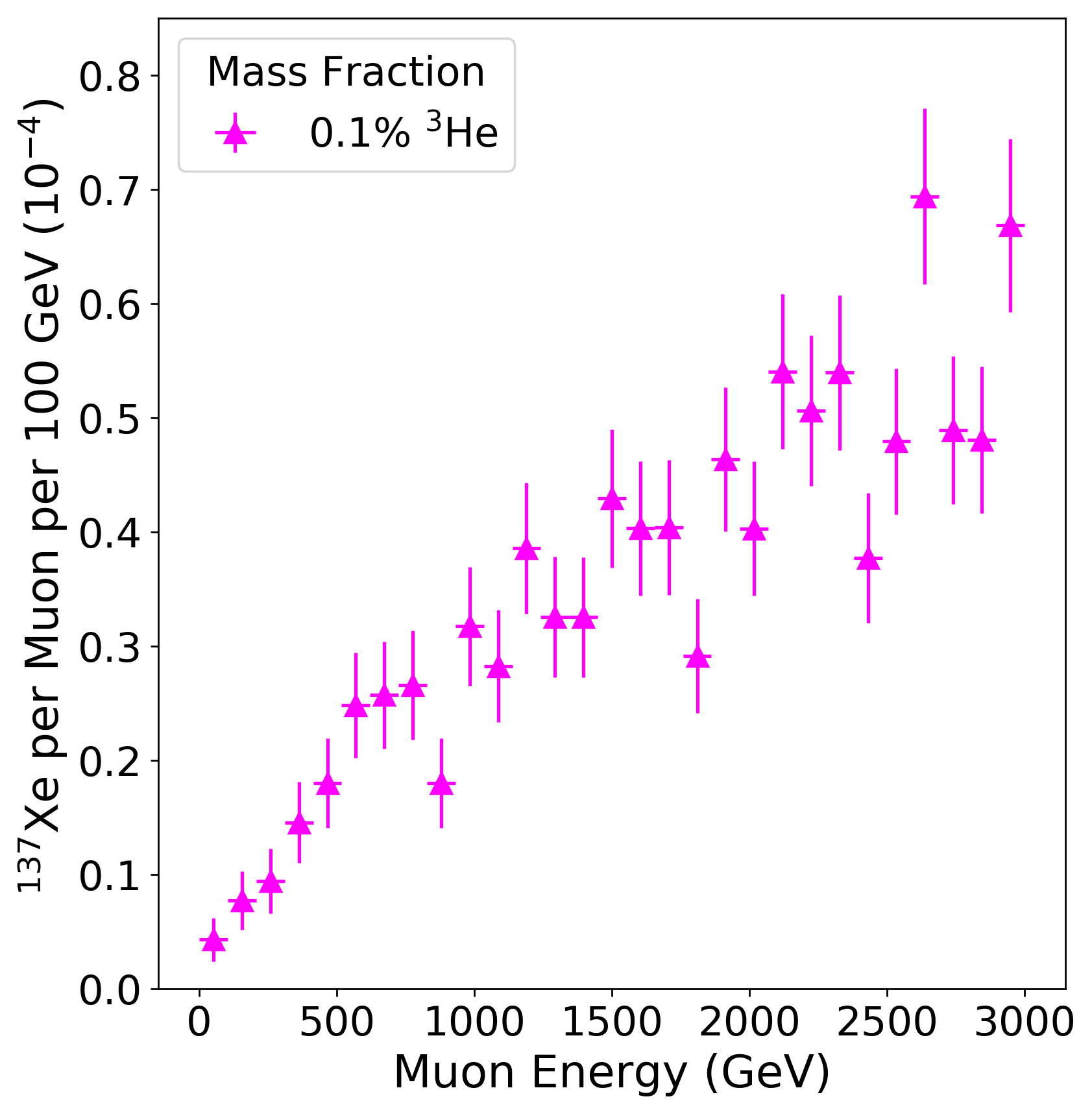}
\includegraphics[width=0.34\linewidth]{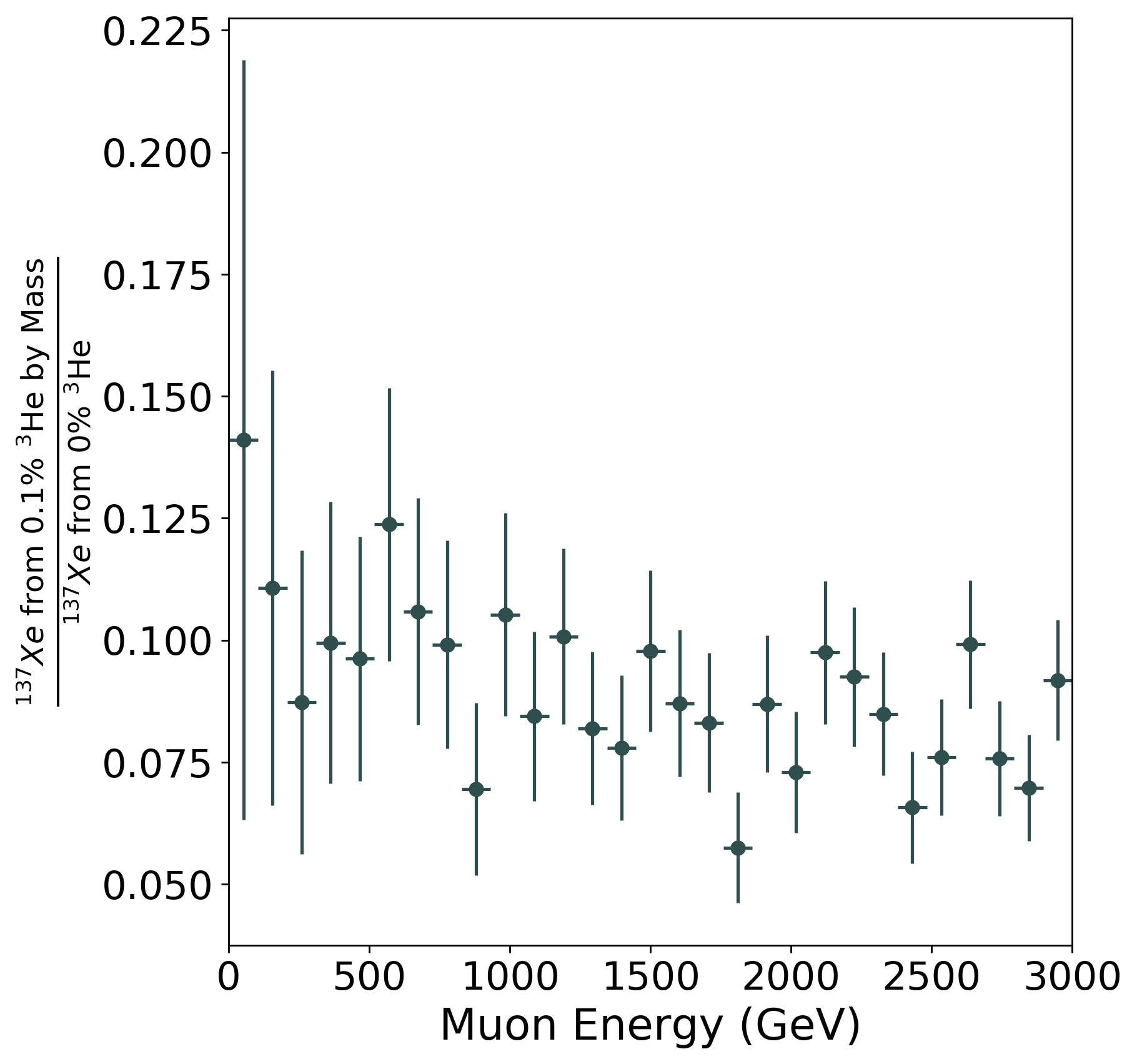}
\caption{Production of \Xe{137} by various cosmic muons energies. \emph{Left}: Production per generated muon for pure enriched xenon; \emph{Center}: Production with \He{3} doped gas. Note that the axes are scaled differently between both plots for easier comparison. \emph{Right}: Ratio of \Xe{137} produced in 0.1\% Helium gas mixture over the production in pure xenon.}
\label{fig:Xe137frommuons}
\end{figure}

The production of \Xe{137} per muon in the detector is the central result required to predict the contamination from \Xe{137} per unit time in a NEXT detector. Figure~\ref{fig:Xe137frommuons} shows the \Xe{137} production expectation per muon in bins of primary muon energy for pure xenon (left) and xenon with $^3$He admixtures (center). As suggested by the neutron production studies, there are more \Xe{137} produced at higher muon energies.  Comparing the left and center plots from Fig.~\ref{fig:Xe137frommuons} (note the different vertical scales) it can be seen that there is significant magnitude reduction in \Xe{137} production with 0.1 percent \He{3} added across all muon energies. The right plot in Fig.~\ref{fig:Xe137frommuons} gives the ratio of the two, showing a cosmogenic background reduction of more than 10 times for 0.1\% of \He{3}.

Additional contributions from neutrons produced by the muons in the rock surrounding the laboratory could be an additional contribution to the activations. Using the same simulation as described above and neutrons starting outside the water tank with energies over the range indicated by figure \ref{fig:NuandMuenergies} we find a \Xe{137} production rate reduction in the presence of 0.1\% \He{3} of the same order as that for muons.

\begin{figure}[t]
\centering
 \includegraphics[height=0.43\linewidth]{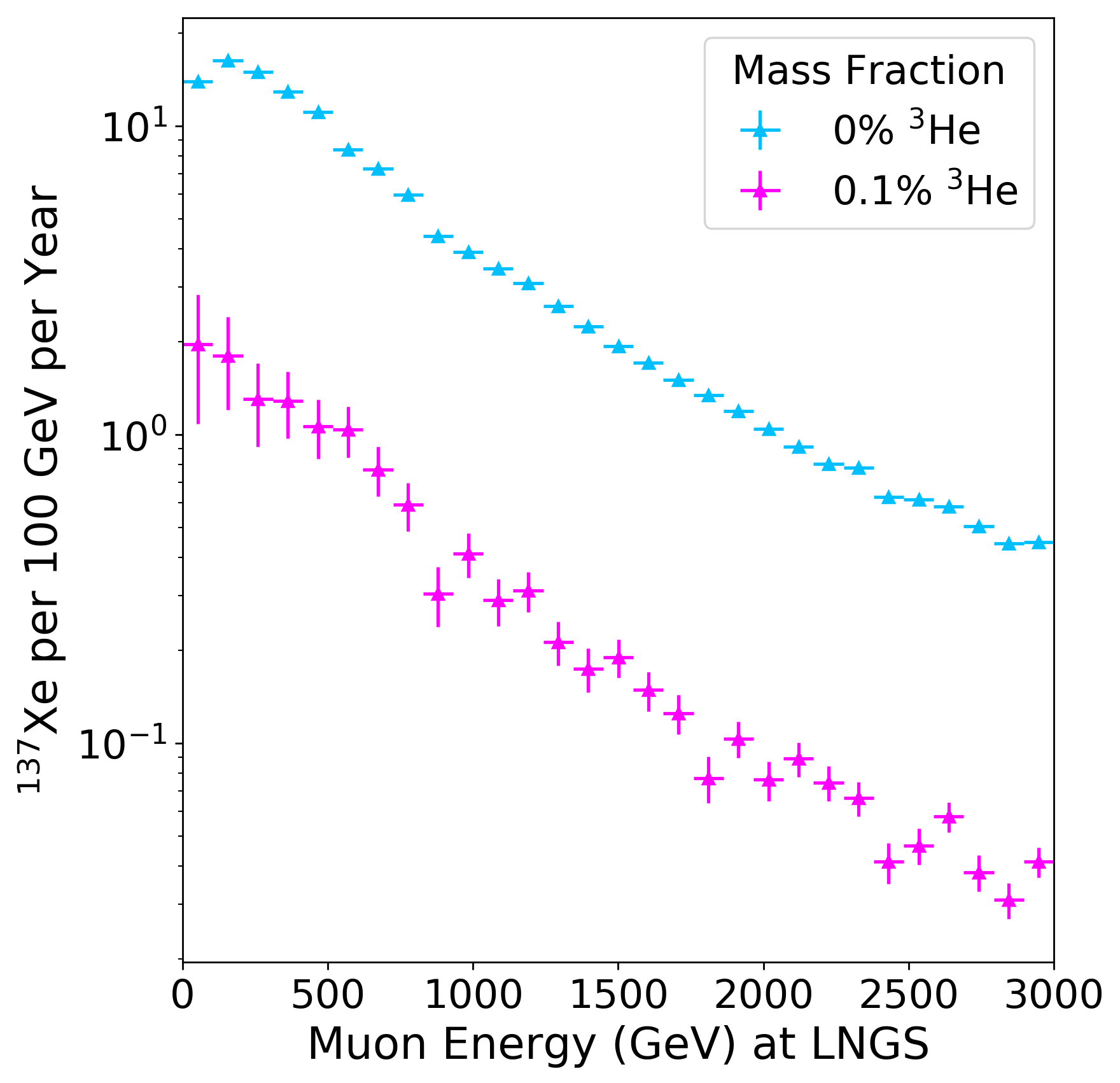}
\includegraphics[height=0.43\linewidth]{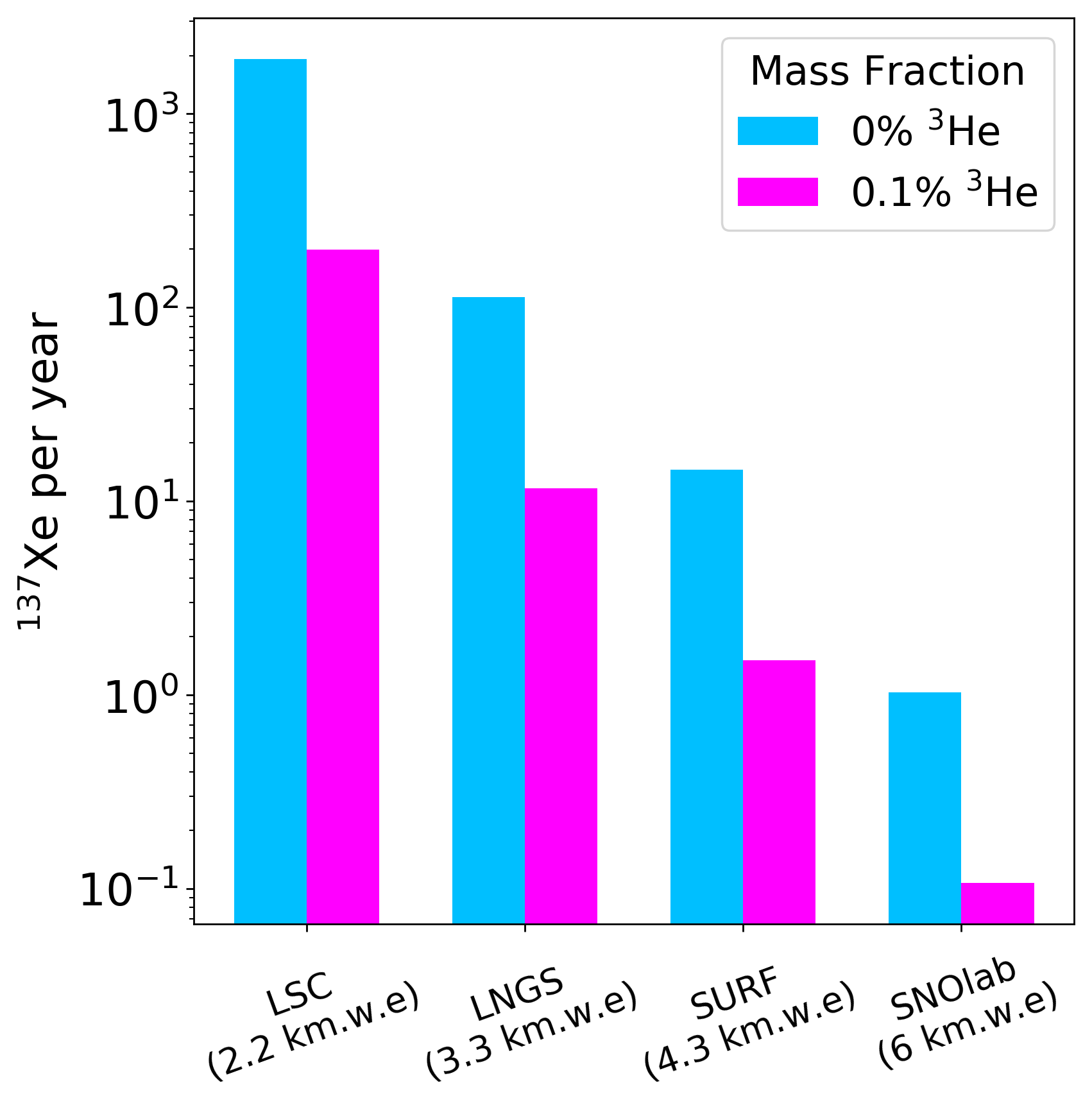}
\caption{\emph{Left}: Rate of \Xe{137} activation expected as a result of the flux distribution of muon energies from LNGS. \emph{Right}: \Xe{137} expected per year in four different  underground laboratory locations and depths. }
\label{fig:labreduction}
\end{figure}

We now consider an example experiment in Laboratori Nazionali del Gran Sasso (LNGS) where the cosmic muon energy spectrum is expected to be that shown in Fig.~\ref{fig:SimGeom}-\emph{left} with an absolute flux normalization of $3.432\times 10^{-8}~\mbox{cm}^{-2}~\mbox{s}^{-1}$ as measured by the Borexino experiment \cite{Agostini:2018fnx}.  The impact on the background index of NEXT can be predicted using the spectrum shown in Fig.~\ref{fig:SimGeom}-\emph{left} convolved with the activation expectations from Fig.~\ref{fig:Xe137frommuons}.  We predict the rate of activations per year shown in Fig.~\ref{fig:labreduction}-\emph{left} for the same experiment with and without 0.1\% \He{3}.

Repeating this exercise for various laboratories (LSC, LNGS, SURF and SNOLab) we find the absolute counts of ~\Xe{137}~yr$^{-1}$ shown in Fig.~\ref{fig:labreduction}-\emph{right}. Considering the full \Xe{137} decay spectrum without any filtering due to analysis, these numbers translate to the activation expectation indices shown in Table~\ref{table:backgrounds}. We note that while such absolute predictions carry a substantial uncertainty due to the physics of neutron production in high energy cascades, our primary results relating to the reduction of activation through \He{3} admixtures are expected to be robust at the twenty percent level. 

To appreciate the impact of these activation rates on a given detector we must account for the acceptance factor for \Xe{137} decay electrons in the signal region of interest, after analysis cuts. Purely energy based arguments can already be used to reject most of these decays as the decay spectrum is broad and \bbnonu\ experiments strive to achieve energy resolution at the few percent level. Most modern xenon gas experiments also have some power to reject single electron events in favour of the double electrons indicative of signal which further reduces this background. If we take into account the topological analysis and energy resolution of the NEXT experiment \cite{Ferrario:2019kwg,Renner:2019pfe}, and a conservative, cut-based analysis, acceptance of \Xe{137} electrons into the signal sample is of order $1.65\times 10^{-4}$ \cite{Javi:thesis} for a symmetric ROI of width 22~keV at \Qbb. Table~\ref{table:backgrounds} also shows the background index that would be expected under these conditions.

\begin{table}[h!]
    \centering
    \begin{tabular}{|c|| c || c | c || c |}\hline
    \ & \multicolumn{2}{c|}{Activation rate} & \multicolumn{2}{c|}{Background index} \\
    \cline{2-5}
    \ & 0\% \He{3}  & 0.1\% \He{3} & 0\% \He{3}  & 0.1\% \He{3}\\
    \ & [~kg$^{-1}$~yr$^{-1}$]  & [~kg$^{-1}$~yr$^{-1}$] & [\ckky] & [\ckky] \\\hline\hline
    LSC & $1.72 \times 10^{0}$ & $1.79\times 10^{-1}$ & $1.29\times 10^{-5}$ &$1.34 \times 10^{-6}$ \\
    LNGS & $1.02 \times 10^{-1}$ & $1.06 \times 10^{-2}$ & $7.65\times 10^{-7}$ &$7.91\times 10^{-8}$ \\   
    SURF & $1.31 \times 10^{-2}$ &$1.36\times 10^{-3}$  & $9.83\times 10^{-8}$ &$1.02\times 10^{-8}$  \\
    SNOlab & $9.29 \times 10^{-4}$ &$9.65\times 10^{-5}$ & $6.97\times 10^{-9}$ &$7.24\times 10^{-10}$ \\\hline 
    \end{tabular}
    \caption{\Xe{137} Activation rate expectations with various percents of helium 3 by mass  and example background indices given an analysis described in the text.}

    \label{table:backgrounds}
\end{table}

 A detailed evaluation of radiogenic backgrounds for the tonne-scale NEXT-HD detector, and how they relate to the initial \Xe{137} background contribution estimated in Tab.1 right, is still underway. However, early estimates suggest that a successful experiment at a relatively shallow location such as LSC would benefit from the addition of \He{3} to the gas. At the multi-tonne scale, the background from  \Xe{137} activation will become truly limiting, and its mitigation via this approach or others may become even more critical.

\section{Economic viability}
Two facts are widely known about \He{3} that should not be left unaddressed: 1) that is it is expensive, and 2) that the supply is limited.  These factors influence discussions of the plausibility of, for example, practical nuclear fusion power based on \He{3}~\cite{wittenberg1986lunar}, which would speculatively consume tens of tonnes of raw \He{3} per year to meet the power needs of the United States.  Such quantities do not exist worldwide at the present time, and have given rise to discussions of exotic acquisition strategies, such as mining the moon~\cite{bilder2009legal,taylor1994helium}, and more realistically in the near-term, breeding in nuclear reactors or extraction from oil and gas reservoirs~\cite{isotope2016defining}.  

Thankfully, far less \He{3} is required to mount a tonne or multi-tonne scale neutrinoless double beta decay program using a \Xe{136} / \He{3} mixture -- on the order of 7500 liters per tonne of \Xe{136} to achieve a 0.1\% by mass concentration.  The use of such quantities of \He{3} is precedented in particle physics instrumentation.  In the 1990s, for example, the SNO~\cite{Ahmad:2002jz} experiment deployed an array of \He{3} counters to detect neutrons produced in neutrino interactions~\cite{amsbaugh2007array}.  The quantity of \He{3} used was approximately 6000~liters~\cite{SnoTritium}, similar to the quantity required for the presented application.  

During the the intervening decades since the existence-proof of the SNO+ \He{3} phase, the economics of \He{3} have changed in important ways.  According to the 2010 Congressional Research Service report {\tt The Helium-3 Shortage: Supply, Demand, and Options for Congress}~\cite{shea2010helium}, ``Helium-3 does not trade in the marketplace as many materials do. It is produced as a byproduct of nuclear weapons maintenance and, in the United States, is then accumulated in a stockpile from which supplies are either transferred directly to other agencies or sold publicly at auction.'' US production in 2015 was estimated to generate approximately 8,000 liters of new \He{3} per year~\cite{isotope2016defining}.  Until 2001 the price at auction was steady at \$100 per liter, a little higher than the per-liter price of \Xe{136}.  However, shortages instigated by the US need for neutron detectors for national security applications after the September 11 attacks of 2001, and the increased use of \He{3} in medical imaging~\cite{tastevin2000optically} led to price spikes, reaching \$2000 / liter at times~\cite{fain2010imaging}.  Even at the highest recent trading prices, however, the cost of the 0.1\% component of \He{3} would be less than that of the 99.9\% \Xe{136} component of the gas mixture.  The stockpile and supply of US \He{3} is now directly controlled by the US Department of Energy, and not traded on an open market.  

While this application would represent a significant fraction of one year's production, \He{3} can be efficiently extracted from \Xe{136} as needed, by liquefying or freezing the xenon and pumping to remove the helium component.  This protocol is commonly employed in experimental studies with Xe/\He{4} mixtures (e.g. Ref.~\cite{McDonald:2019fhy}).  The separation process can be performed either completely or partially, as need arises, and so this application would represent storage and stewardship, rather than irreversible consumption.   Furthermore, although \He{3} is a limited and expensive resource, it is notable that the majority isotope in this mixture is \Xe{136}, the world production and stockpile of which would both be zero, were it not for neutrinoless double beta decay experiments.  Thus the difficulties associated with acquisition of the minority \He{3} component for temporary use in this manner should be assessed in relative terms.  They do not appear prohibitive, based on the last traded market price, current levels of production, and precedented use cases in particle physics.

We may also consider alternative gases that have been explored to play a similar role to \He{3} in the face of limited supply.  BF$_3$ enriched in $^{10}$B is one attractive possibility~\cite{korff1939neutron,lintereur20113he,goodings1972neutron,fowler1950boron,segre1947boron}. This gas has a high neutron capture cross section, can be mixed into xenon, and should not attach electrons.  The challenge associated with BF$_3$ is that it is toxic, making operation of a detector in an underground laboratory challenging from a safety perspective.  On the other hand, the \Xe{136} used in double beta decay experiments is sufficiently precious that they typically have elaborate systems to recapture the gas and to minimizes losses, even in small quantities, so perhaps BF$_3$ should not be immediately dismissed. Though this gas could offer the same function as a \He{3} additive  without supply challenges, the R\&D associated with circulating, purifying, achieving energy resolution in a Xe/BF$_3$ mix is likely more involved than with a Xe/He mix, which is why \He{3} has been our primary focus in this work.

\section{Conclusions}
The impact of the addition of small percentages of \He{3} to a tonne-scale underground high pressure xenon gas detector resembling NEXT-HD has been studied as a means to reduce backgrounds from the capture of thermal neutrons on \Xe{136}. Studies with injected neutrons show a reduction in the number of these activations of over 1 order of magnitude with as little as  $0.1\%$ by mass of \He{3} doping.  For higher energy neutrons, multi-neutron production from one initial parent is present, leading to larger production of \Xe{137} per primary neutron.

An example experiment with a fiducial mass of approximately 1~tonne surrounded with a water tank was used to study the impact on the background induced by the passage of cosmic muons. If the experiment took data at LNGS, an activation rate of $1.02 \times 10^{-1}$~kg$^{-1}$~yr$^{-1}$ would be expected in the case of pure enriched xenon operation with a reduction to $1.24 \times 10^{-2}$~kg$^{-1}$~yr$^{-1}$ expected with the addition of 0.1\% \He{3} by mass. Similar predictions have been made for other underground sites.  In addition to reducing background, the observed suppression of \Xe{137} can be used to relax requirements on the outer shielding, and potentially loosen analysis cuts designed to filter backgrounds from \Xe{137} for enhanced signal acceptance. 

Given the background reduction power of high pressure xenon gas TPCs against beta decays from \Xe{137}, it is expected that any moderately deep underground laboratory, a \Xe{136}/\He{3} tonne or multi-tonne-scale experiment will be entirely free of background from cosmogenically activated \Xe{137}.   As shown in Fig.~\ref{fig:labreduction}, with \Xe{136}/\He{3} the number of \Xe{137} expected per year in LSC, the present home of NEXT-100 and the least deep laboratory considered, is lower than levels that would have a substantial effect on experimental sensitivity. This is true even when accounting for  sizeable uncertainties arising from cosmogenic shower modelling.  We conclude that \Xe{136}/\He{3} mixtures may represent a promising technological component for future large high pressure xenon gas experiments.

\section*{Acknowledgements}
The work described was supported by the Department of Energy under Award numbers {DE-SC0019054} and {DE-SC0019223}. The NEXT Collaboration acknowledges support from the following agencies and institutions: the European Research Council (ERC) under the Advanced Grant 339787-NEXT; the European Union's Framework Programme for Research and Innovation Horizon 2020 (2014-2020) under the Marie Sk\l odowska-Curie Grant Agreements No. 674896, 690575 and 740055; the Ministerio de Econom\'ia y Competitividad of Spain under grants FIS2014-53371-C04, the Severo Ochoa Program SEV-2014-0398 and the Mar\'ia de Maetzu Program MDM-2016-0692; the GVA of Spain under grants PROMETEO/2016/120 and SEJI/2017/011; the Portuguese FCT under project PTDC/FIS-NUC/2525/2014, under project UID/FIS/04559/2013 to fund the activities of LIBPhys, and under grants PD/BD/105921/2014, SFRH/BPD/109180/2015 and SFRH/BPD/76842/2011. Finally, we are grateful to the Laboratorio Subterr\'aneo de Canfranc for hosting and supporting the NEXT experiment.

\bibliographystyle{JHEP}
\bibliography{main}
\begin{appendices}
\label{sec:Checks}
During the writing of this document a newer version of {\tt GEANT4} was produced ({\tt 10.6.p01}) with some updates to the neutron production cross section. As a general cross check of the results additional data points were produced using this version of the simulation code.

\begin{figure}[b!]
\centering
 \includegraphics[height=0.43\linewidth]{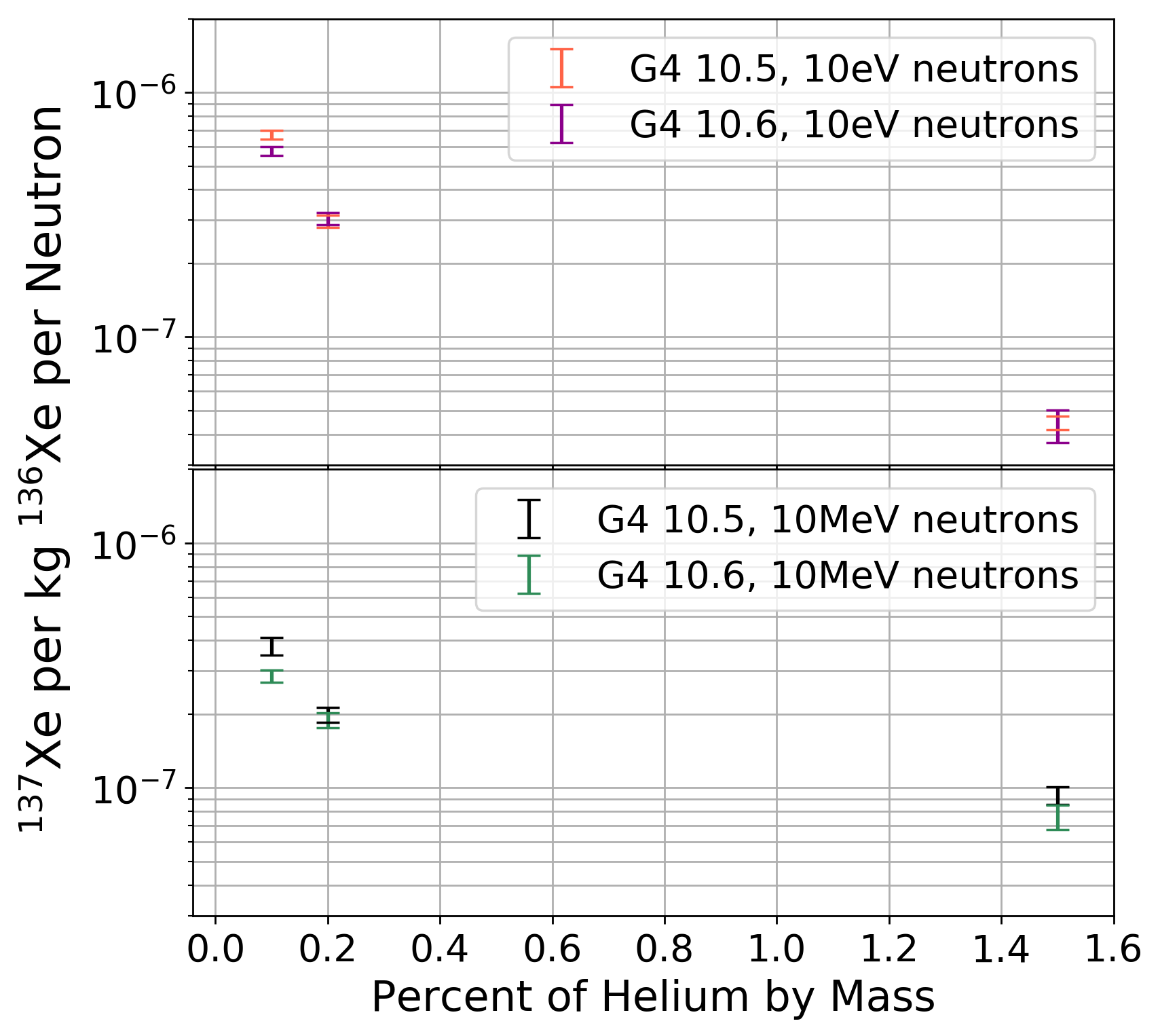}
\includegraphics[height=0.43\linewidth]{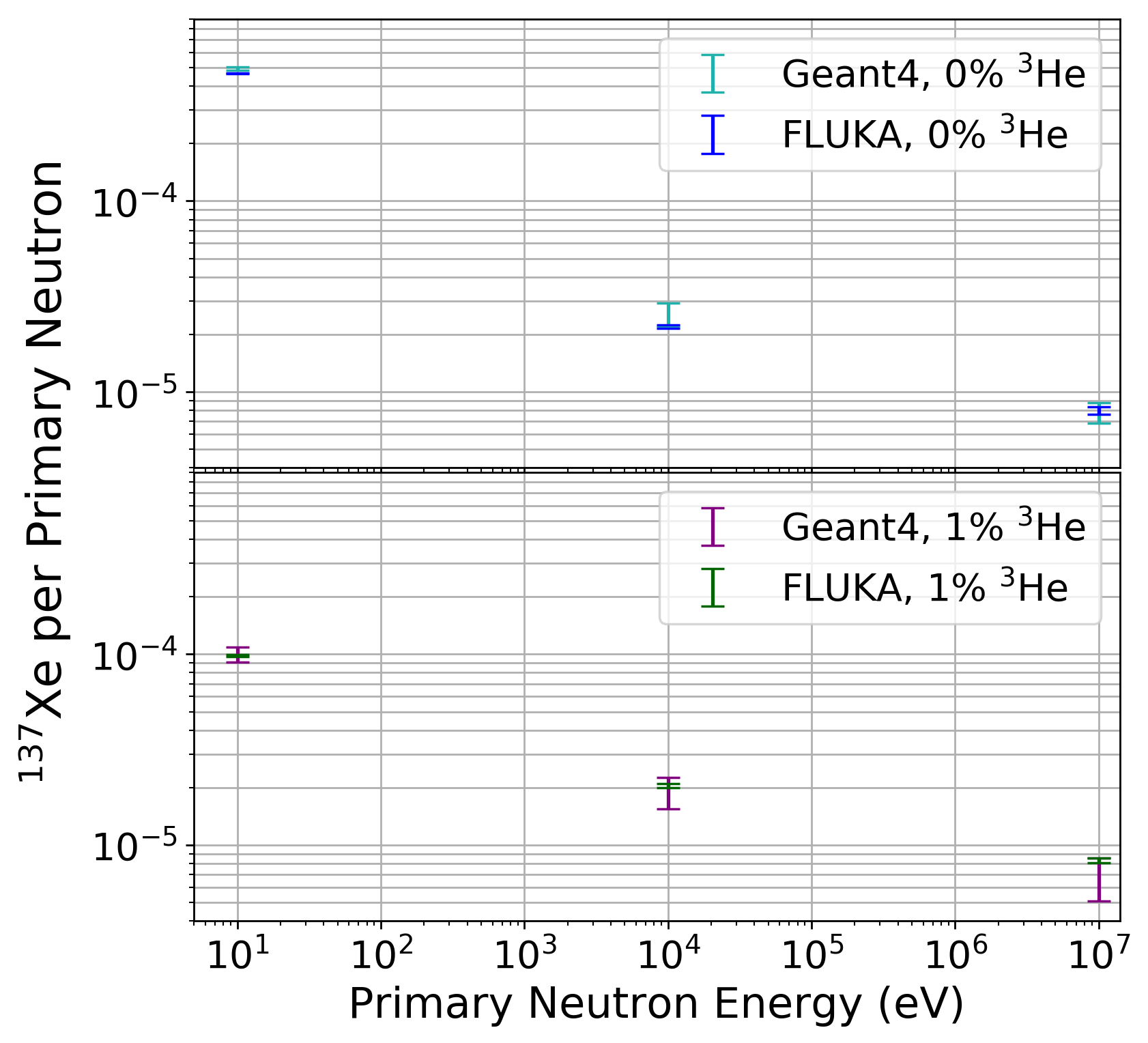}
\caption{\emph{Left}: Rate of \Xe{137} activation expected as a result of thermal and fast neutrons with different {\tt GEANT4} versions. \emph{Right}: \Xe{137} activation rates using {\tt FLUKA} and {\tt GEANT4} {\tt 10.5.p01.}}
\label{fig:checks}
\end{figure}

Fast (10~MeV) and thermal (10~eV) neutrons were passed through the geometry as described in Section~\ref{sec:NEXT} and the production rates for \Xe{137} can be seen in Fig. \ref{fig:checks}-\emph{left}. We observed some systematic difference  between {\tt GEANT4} versions ({\tt 10.6.p01}) and ({\tt 10.5.p01}), with the largest discrepancy being for fast neutrons at 0.1\% helium, where the ratio between simulated capture rates was 1.329 $\pm$ 0.219.  This is within the envelope of systematic uncertainty ascribed to our results, and does not affect our primary conclusions.

As a further check of the validity of the results simulation of neutrons in {\tt GEANT4.10.5.p01} was compared to that in the {\tt 2011.2x-8} {\tt FLUKA} MonteCarlo simulation package~\cite{BOHLEN2014211}~\cite{Ferrari:2005zk}: a widely trusted tool for transport and interactions of low-energy neutrons.
A simple 1~m by 1~m cylinder of enriched xenon gas and a mono-energetic beam of neutrons were simulated using both packages. Neutrons were injected uniformly along, and perpendicular to, one endcap of the cylinder. Fig.~\ref{fig:checks}-\emph{right} shows a comparison of the rate of production of \Xe{137} in each simulation for a range of neutron energies. The results of {\tt GEANT4.10.5.p01} are compatible with {\tt 2011.2x-8} {\tt FLUKA} within systematic uncertainties.

\end{appendices} 

\end{document}

%% file: src/Commands.tex
\newcommand{\He}[1]{$^{#1}$He}
\newcommand{\Xe}[1]{$^{#1}$Xe}
\newcommand{\Bi}[1]{$^{#1}$Bi}
\newcommand{\Tl}[1]{$^{#1}$Tl}
\newcommand{\bbnonu}{$0\nu\beta\beta$}

\newcommand{\Qbb}{$Q_{\beta\beta}$}

\newcommand{\Qb}{$Q_{\beta}$}

\newcommand{\ckky}{keV$^{-1}$kg$^{-1}$yr$^{-1}$}

%% file: src/Authors.tex
\collaboration{The NEXT Collaboration}
\author[3,a]{L.~Rogers,\note[a]{Corresponding author.}}
\author[3]{B.J.P.~Jones,}
\author[3]{A.~Laing,}

\author[3]{S.~Pingulkar,}

\author[3]{B.~Smithers,}
\author[3]{K.~Woodruff,}
\author[2]{C.~Adams,}
\author[22]{V.~\'Alvarez,}
\author[6]{L.~Arazi,}
\author[20]{I.J.~Arnquist,}
\author[4]{C.D.R~Azevedo,}
\author[2]{K.~Bailey,}
\author[22]{F.~Ballester,}
\author[19]{J.M.~Benlloch-Rodr\'{i}guez,}
\author[14]{F.I.G.M.~Borges,}
\author[3]{N.~Byrnes,}
\author[19]{S.~C\'arcel,}
\author[19]{J.V.~Carri\'on,}
\author[23]{S.~Cebri\'an,}
\author[20]{E.~Church,}
\author[14]{C.A.N.~Conde,}
\author[11]{T.~Contreras,}
\author[21]{G.~D\'iaz,}
\author[19]{J.~D\'iaz,}
\author[5]{M.~Diesburg,}
\author[3]{R.~Dingler,}
\author[14]{J.~Escada,}
\author[22]{R.~Esteve,}
\author[6,19]{R.~Felkai,}
\author[13]{A.F.M.~Fernandes,}
\author[13]{L.M.P.~Fernandes,}
\author[16,9]{P.~Ferrario,}
\author[4]{A.L.~Ferreira,}
\author[13]{E.D.C.~Freitas,}
\author[16]{J.~Generowicz,}
\author[11]{S.~Ghosh,}
\author[8]{A.~Goldschmidt,}
\author[16,9,b]{J.J.~G\'omez-Cadenas,\note[b]{NEXT Co-spokesperson.}}
\author[21]{D.~Gonz\'alez-D\'iaz,}
\author[11]{R.~Guenette,}
\author[10]{R.M.~Guti\'errez,}
\author[11]{J.~Haefner,}
\author[2]{K.~Hafidi,}
\author[1]{J.~Hauptman,}
\author[13]{C.A.O.~Henriques,}
\author[21]{J.A.~Hernando~Morata,}
\author[16]{P.~Herrero,}
\author[22]{V.~Herrero,}
\author[6,7]{Y.~Ifergan,}
\author[2]{S.~Johnston,}
\author[19,21]{M.~Kekic,}
\author[18]{L.~Labarga,}
\author[5]{P.~Lebrun,}
\author[19]{N.~L\'opez-March,}
\author[10]{M.~Losada,}
\author[13]{R.D.P.~Mano,}
\author[11,8]{J.~Mart\'in-Albo,}
\author[19]{A.~Mart\'inez,}
\author[19,21,c]{G.~Mart\'inez-Lema,\note[c]{Now at Weizmann Institute of Science, Israel.}}
\author[3]{A.D.~McDonald,}
\author[16,9]{F.~Monrabal,}
\author[13]{C.M.B.~Monteiro,}
\author[22]{F.J.~Mora,}
\author[19]{J.~Mu\~noz~Vidal,}
\author[19]{P.~Novella,}
\author[3,b]{D.R.~Nygren,}
\author[19]{B.~Palmeiro,}
\author[5]{A.~Para,}
\author[12]{J.~P\'erez,}
\author[19]{M.~Querol,}
\author[6]{A.B.~Redwine,}
\author[21]{J.~Renner,}
\author[2]{J.~Repond,}
\author[2]{S.~Riordan,}
\author[17]{L.~Ripoll,}
\author[10]{Y.~Rodr\'iguez~Garc\'ia,}
\author[16]{J.~Rodr\'iguez,}
\author[16,12]{B.~Romeo,}
\author[19]{C.~Romo-Luque,}
\author[14]{F.P.~Santos,}
\author[13]{J.M.F.~dos~Santos,}
\author[6]{A.~Sim\'on,}
\author[15,e]{C.~Sofka,\note[e]{Now at University of Texas at Austin, USA.}}
\author[19]{M.~Sorel,}
\author[15]{T.~Stiegler,}
\author[22]{J.F.~Toledo,}
\author[16]{J.~Torrent,}
\author[19]{A.~Us\'on,}
\author[4]{J.F.C.A.~Veloso,}
\author[15]{R.~Webb,}
\author[6,f]{R.~Weiss-Babai,\note[f]{On leave from Soreq Nuclear Research Center, Yavneh, Israel.}}
\author[15,g]{J.T.~White,\note[g]{Deceased.}}
\author[19]{N.~Yahlali}
\emailAdd{leslie.rogers@mavs.uta.edu}
\affiliation[1]{
Department of Physics and Astronomy, Iowa State University, 12 Physics Hall, Ames, IA 50011-3160, USA}
\affiliation[2]{
Argonne National Laboratory, Argonne, IL 60439, USA}
\affiliation[3]{
Department of Physics, University of Texas at Arlington, Arlington, TX 76019, USA}
\affiliation[4]{
Institute of Nanostructures, Nanomodelling and Nanofabrication (i3N), Universidade de Aveiro, Campus de Santiago, Aveiro, 3810-193, Portugal}
\affiliation[5]{
Fermi National Accelerator Laboratory, Batavia, IL 60510, USA}
\affiliation[6]{
Nuclear Engineering Unit, Faculty of Engineering Sciences, Ben-Gurion University of the Negev, P.O.B. 653, Beer-Sheva, 8410501, Israel}
\affiliation[7]{
Nuclear Research Center Negev, Beer-Sheva, 84190, Israel}
\affiliation[8]{
Lawrence Berkeley National Laboratory (LBNL), 1 Cyclotron Road, Berkeley, CA 94720, USA}
\affiliation[9]{
Ikerbasque, Basque Foundation for Science, Bilbao, E-48013, Spain}
\affiliation[10]{
Centro de Investigaci\'on en Ciencias B\'asicas y Aplicadas, Universidad Antonio Nari\~no, Sede Circunvalar, Carretera 3 Este No.\ 47 A-15, Bogot\'a, Colombia}
\affiliation[11]{
Department of Physics, Harvard University, Cambridge, MA 02138, USA}
\affiliation[12]{
Laboratorio Subterr\'aneo de Canfranc, Paseo de los Ayerbe s/n, Canfranc Estaci\'on, E-22880, Spain}
\affiliation[13]{
LIBPhys, Department of Physics, University of Coimbra, Rua Larga, Coimbra, 3004-516, Portugal}
\affiliation[14]{
LIP, Department of Physics, University of Coimbra, Coimbra, 3004-516, Portugal}
\affiliation[15]{
Department of Physics and Astronomy, Texas A\&M University, College Station, TX 77843-4242, USA}
\affiliation[16]{
Donostia International Physics Center (DIPC), Paseo Manuel Lardizabal, 4, Donostia-San Sebastian, E-20018, Spain}
\affiliation[17]{
Escola Polit\`ecnica Superior, Universitat de Girona, Av.~Montilivi, s/n, Girona, E-17071, Spain}
\affiliation[18]{
Departamento de F\'isica Te\'orica, Universidad Aut\'onoma de Madrid, Campus de Cantoblanco, Madrid, E-28049, Spain}
\affiliation[19]{
Instituto de F\'isica Corpuscular (IFIC), CSIC \& Universitat de Val\`encia, Calle Catedr\'atico Jos\'e Beltr\'an, 2, Paterna, E-46980, Spain}
\affiliation[20]{
Pacific Northwest National Laboratory (PNNL), Richland, WA 99352, USA}
\affiliation[21]{
Instituto Gallego de F\'isica de Altas Energ\'ias, Univ.\ de Santiago de Compostela, Campus sur, R\'ua Xos\'e Mar\'ia Su\'arez N\'u\~nez, s/n, Santiago de Compostela, E-15782, Spain}
\affiliation[22]{
Instituto de Instrumentaci\'on para Imagen Molecular (I3M), Centro Mixto CSIC - Universitat Polit\`ecnica de Val\`encia, Camino de Vera s/n, Valencia, E-46022, Spain}
\affiliation[23]{
Centro de Astropart\'iculas y F\'isica de Altas Energ\'ias (CAPA),Universidad de Zaragoza, Calle Pedro Cerbuna, 12, Zaragoza, E-50009, Spain}

%% file: main.bbl
\providecommand{\href}[2]{#2}\begingroup\raggedright\begin{thebibliography}{10}

\bibitem{Nygren:2009zz}
D.~Nygren, {\it {High-pressure xenon gas electroluminescent TPC for $0-\nu ~
  \beta \beta$-decay search}},  {\em Nucl.Instrum.Meth.} {\bf A603} (2009)
  337--348.

\bibitem{Alvarez:2012haa}
{\bf NEXT} Collaboration, V.~\'Alvarez et~al., {\it {NEXT-100 Technical Design
  Report (TDR): Executive Summary}},  {\em JINST} {\bf 7} (2012) T06001,
  [\href{http://xxx.lanl.gov/abs/1202.0721}{{\tt arXiv:1202.0721}}].

\bibitem{Alvarez:2013gxa}
{\bf NEXT} Collaboration, V.~Alvarez et~al., {\it {Operation and first results
  of the NEXT-DEMO prototype using a silicon photomultiplier tracking array}},
  {\em JINST} {\bf 8} (2013) P09011,
  [\href{http://xxx.lanl.gov/abs/1306.0471}{{\tt arXiv:1306.0471}}].

\bibitem{Ferrario:2015kta}
{\bf NEXT} Collaboration, P.~Ferrario et~al., {\it {First proof of topological
  signature in the high pressure xenon gas TPC with electroluminescence
  amplification for the NEXT experiment}},  {\em JHEP} {\bf 01} (2016) 104,
  [\href{http://xxx.lanl.gov/abs/1507.05902}{{\tt arXiv:1507.05902}}].

\bibitem{Monrabal:2018xlr}
{\bf NEXT} Collaboration, F.~Monrabal et~al., {\it {The Next White (NEW)
  Detector}},  {\em JINST} {\bf 13} (2018), no.~12 P12010,
  [\href{http://xxx.lanl.gov/abs/1804.02409}{{\tt arXiv:1804.02409}}].

\bibitem{Ferrario:2019kwg}
{\bf NEXT} Collaboration, P.~Ferrario et~al., {\it {Demonstration of the event
  identification capabilities of the NEXT-White detector}},  {\em JHEP} {\bf
  10} (2019) 052, [\href{http://xxx.lanl.gov/abs/1905.13141}{{\tt
  arXiv:1905.13141}}].

\bibitem{Renner:2019pfe}
{\bf NEXT} Collaboration, J.~Renner et~al., {\it {Energy calibration of the
  NEXT-White detector with 1\% resolution near Q$_{\beta\beta}$ of
  $^{136}$Xe}},  {\em JHEP} {\bf 10} (2019) 230,
  [\href{http://xxx.lanl.gov/abs/1905.13110}{{\tt arXiv:1905.13110}}].

\bibitem{Novella:2019cne}
{\bf NEXT} Collaboration, P.~Novella et~al., {\it {Radiogenic Backgrounds in
  the NEXT Double Beta Decay Experiment}},  {\em JHEP} {\bf 10} (2019) 051,
  [\href{http://xxx.lanl.gov/abs/1905.13625}{{\tt arXiv:1905.13625}}].

\bibitem{Martin-Albo:2015rhw}
{\bf NEXT} Collaboration, J.~Martin-Albo et~al., {\it {Sensitivity of NEXT-100
  to Neutrinoless Double Beta Decay}},  {\em JHEP} {\bf 05} (2016) 159,
  [\href{http://xxx.lanl.gov/abs/1511.09246}{{\tt arXiv:1511.09246}}].

\bibitem{XePa}
B.~J.~P. Jones, ``{XePA Project: Drift properties of helium added to xenon at
  10 bar}.'' \url{http://www-hep.uta.edu/~bjones/XePA/}, 2016.

\bibitem{Felkai:2017oeq}
R.~Felkai et~al., {\it {Helium-Xenon mixtures to improve the topological
  signature in high pressure gas xenon TPCs}},  {\em Nucl. Instrum. Meth.} {\bf
  A905} (2018) 82--90, [\href{http://xxx.lanl.gov/abs/1710.05600}{{\tt
  arXiv:1710.05600}}].

\bibitem{Henriques:2018tam}
{\bf NEXT} Collaboration, C.~A.~O. Henriques et~al., {\it {Electroluminescence
  TPCs at the Thermal Diffusion Limit}},  {\em JHEP} {\bf 01} (2019) 027,
  [\href{http://xxx.lanl.gov/abs/1806.05891}{{\tt arXiv:1806.05891}}].

\bibitem{McDonald:2019fhy}
{\bf NEXT} Collaboration, A.~D. McDonald et~al., {\it {Electron Drift and
  Longitudinal Diffusion in High Pressure Xenon-Helium Gas Mixtures}},  {\em
  JINST} {\bf 14} (2019), no.~08 P08009,
  [\href{http://xxx.lanl.gov/abs/1902.05544}{{\tt arXiv:1902.05544}}].

\bibitem{Fernandes:2019zuz}
{\bf NEXT} Collaboration, A.~F.~M. Fernandes et~al., {\it {Electroluminescence
  Yield in low-diffusion Xe-He gas mixtures}},
  \href{http://xxx.lanl.gov/abs/1906.03984}{{\tt arXiv:1906.03984}}.

\bibitem{Lippincott:2017yst}
H.~Lippincott, T.~Alexander, and A.~Hime, {\it {Increasing the sensitivity of
  LXe TPCs to dark matter by doping with helium or neon}},  {\em PoS} {\bf
  ICHEP2016} (2017) 285.

\bibitem{Anton:2019wmi}
{\bf EXO-200} Collaboration, G.~Anton et~al., {\it {Search for Neutrinoless
  Double-Beta Decay with the Complete EXO-200 Dataset}},  {\em Phys. Rev.
  Lett.} {\bf 123} (2019), no.~16 161802,
  [\href{http://xxx.lanl.gov/abs/1906.02723}{{\tt arXiv:1906.02723}}].

\bibitem{EXO200::2015wtc}
{\bf EXO-200} Collaboration, J.~B. Albert et~al., {\it {Cosmogenic Backgrounds
  to $0{\nu}{\beta}{\beta}$ in EXO-200}},  {\em JCAP} {\bf 1604} (2016), no.~04
  029, [\href{http://xxx.lanl.gov/abs/1512.06835}{{\tt arXiv:1512.06835}}].

\bibitem{Albert:2017hjq}
{\bf nEXO} Collaboration, J.~B. Albert et~al., {\it {Sensitivity and Discovery
  Potential of nEXO to Neutrinoless Double Beta Decay}},  {\em Phys. Rev.} {\bf
  C97} (2018), no.~6 065503, [\href{http://xxx.lanl.gov/abs/1710.05075}{{\tt
  arXiv:1710.05075}}].

\bibitem{Gando_2016}
A.~Gando, Y.~Gando, T.~Hachiya, A.~Hayashi, S.~Hayashida, H.~Ikeda, K.~Inoue,
  K.~Ishidoshiro, Y.~Karino, M.~Koga, and et~al., {\it Search for majorana
  neutrinos near the inverted mass hierarchy region with kamland-zen},  {\em
  Physical Review Letters} {\bf 117} (Aug, 2016).

\bibitem{jones2016single}
B.~Jones, A.~McDonald, and D.~Nygren, {\it Single molecule fluorescence imaging
  as a technique for barium tagging in neutrinoless double beta decay},  {\em
  Journal of Instrumentation} {\bf 11} (2016), no.~12 P12011.

\bibitem{mcdonald2018demonstration}
A.~McDonald, B.~Jones, D.~Nygren, C.~Adams, V.~{\'A}lvarez, C.~Azevedo,
  J.~Benlloch-Rodr{\'\i}guez, F.~Borges, A.~Botas, S.~C{\'a}rcel, et~al., {\it
  Demonstration of single-barium-ion sensitivity for neutrinoless double-beta
  decay using single-molecule fluorescence imaging},  {\em Physical review
  letters} {\bf 120} (2018), no.~13 132504.

\bibitem{byrnes2019barium}
N.~Byrnes, A.~Denisenko, F.~Foss~Jr, B.~Jones, A.~McDonald, D.~Nygren,
  P.~Thapa, and K.~Woodruff, {\it Barium tagging with selective,
  dry-functional, single molecule sensitive on-off fluorophores for the next
  experiment},  {\em arXiv preprint arXiv:1909.04677} (2019).

\bibitem{Thapa:2019zjk}
P.~Thapa, I.~Arnquist, N.~Byrnes, A.~A. Denisenko, F.~W. Foss, B.~J.~P. Jones,
  A.~D. Mcdonald, D.~R. Nygren, and K.~Woodruff, {\it {Barium Chemosensors with
  Dry-Phase Fluorescence for Neutrinoless Double Beta Decay}},  {\em Sci. Rep.}
  {\bf 9} (2019), no.~1 15097, [\href{http://xxx.lanl.gov/abs/1904.05901}{{\tt
  arXiv:1904.05901}}].

\bibitem{Rivilla:2019vzd}
I.~Rivilla et~al., {\it {Towards a background-free neutrinoless double beta
  decay experiment based on a fluorescent bicolor sensor}},
  \href{http://xxx.lanl.gov/abs/1909.02782}{{\tt arXiv:1909.02782}}.

\bibitem{Woodruff:2019hte}
K.~Woodruff et~al., {\it {Radio Frequency and DC High Voltage Breakdown of High
  Pressure Helium, Argon, and Xenon}},
  \href{http://xxx.lanl.gov/abs/1909.05860}{{\tt arXiv:1909.05860}}.

\bibitem{chadwick2011endf}
M.~Chadwick, M.~Herman, P.~Oblo{\v{z}}insk{\`y}, M.~E. Dunn, Y.~Danon,
  A.~Kahler, D.~L. Smith, B.~Pritychenko, G.~Arbanas, R.~Arcilla, et~al., {\it
  Endf/b-vii. 1 nuclear data for science and technology: cross sections,
  covariances, fission product yields and decay data},  {\em Nuclear data
  sheets} {\bf 112} (2011), no.~12 2887--2996.

\bibitem{Brown:2018jhj}
D.~A. Brown et~al., {\it {ENDF/B-VIII.0: The 8th Major Release of the Nuclear
  Reaction Data Library with CIELO-project Cross Sections, New Standards and
  Thermal Scattering Data}},  {\em Nucl. Data Sheets} {\bf 148} (2018) 1--142.

\bibitem{Martinez-Lema:2018ibw}
{\bf NEXT} Collaboration, G.~Martinez-Lema et~al., {\it {Calibration of the
  NEXT-White detector using $^{83m}\mathrm{Kr}$ decays}},  {\em JINST} {\bf 13}
  (2018), no.~10 P10014, [\href{http://xxx.lanl.gov/abs/1804.01780}{{\tt
  arXiv:1804.01780}}].

\bibitem{SnoTritium}
S.~R. Elliott, K.~M. Heeger, A.~Hime, R.~G.~H. Robertson, M.~W.~E. Smith, and
  T.~D. Steiger, {\it Tritium contamination in ncds}, .
  \url{https://sno.phy.queensu.ca/sno/str/SNO-STR-99-006.ps.gz}.

\bibitem{Agostinelli:2002hh}
{\bf GEANT4} Collaboration, S.~Agostinelli et~al., {\it {GEANT4: A Simulation
  toolkit}},  {\em Nucl. Instrum. Meth.} {\bf A506} (2003) 250--303.

\bibitem{mughabghab2006atlas}
S.~F. Mughabghab, {\em Atlas of Neutron Resonances: Resonance Parameters and
  Thermal Cross Sections. Z= 1-100}.
\newblock Elsevier, 2006.

\bibitem{Albert:2016vmb}
J.~B. Albert, S.~J. Daugherty, T.~N. Johnson, T.~O'Conner, L.~Kaufman,
  A.~Couture, J.~L. Ullmann, and M.~Krti\v{c}ka, {\it {Measurement of neutron
  capture on $^{136}$Xe}},  {\em Phys. Rev.} {\bf C94} (2016), no.~3 034617,
  [\href{http://xxx.lanl.gov/abs/1605.05794}{{\tt arXiv:1605.05794}}].

\bibitem{batchelor1955helium}
R.~Batchelor, R.~Aves, and T.~Skyrme, {\it Helium-3 filled proportional counter
  for neutron spectroscopy},  {\em Review of Scientific Instruments} {\bf 26}
  (1955), no.~11 1037--1047.

\bibitem{manokhin1988brond}
V.~Manokhin, {\it Brond: Ussr recommended evaluated neutron data library},
  tech. rep., International Atomic Energy Agency, 1988.

\bibitem{gibbons1959total}
J.~Gibbons and R.~Macklin, {\it Total neutron yields from light elements under
  proton and alpha bombardment},  {\em Physical Review} {\bf 114} (1959), no.~2
  571.

\bibitem{haesner1983measurement}
B.~Haesner, W.~Heeringa, H.~Klages, H.~Dobiasch, G.~Schmalz, P.~Schwarz,
  J.~Wilczynski, B.~Zeitnitz, and F.~K{\"a}ppeler, {\it Measurement of the he 3
  and he 4 total neutron cross sections up to 40 mev},  {\em Physical Review C}
  {\bf 28} (1983), no.~3 995.

\bibitem{antolkovic1967study}
B.~Antolkovi{\'c}, G.~Pai{\'c}, P.~Toma{\v{s}}, and D.~Rendi{\'c}, {\it Study
  of neutron-induced reactions on he 3 at e n= 14.4 mev},  {\em Physical
  Review} {\bf 159} (1967), no.~4 777.

\bibitem{seagrave1960elastic}
J.~Seagrave, L.~Cranberg, and J.~Simmons, {\it Elastic scattering of fast
  neutrons by tritium and he 3},  {\em Physical Review} {\bf 119} (1960), no.~6
  1981.

\bibitem{sayres1961interaction}
A.~R. Sayres, K.~Jones, and C.~Wu, {\it Interaction of neutrons with he 3},
  {\em Physical Review} {\bf 122} (1961), no.~6 1853.

\bibitem{als1964slow}
J.~Als-Nielsen and O.~Dietrich, {\it Slow neutron cross sections for he 3, b,
  and au},  {\em Physical Review} {\bf 133} (1964), no.~4B B925.

\bibitem{Bertini:1963zzc}
H.~W. Bertini, {\it {Low-Energy Intranuclear Cascade Calculation}},  {\em Phys.
  Rev.} {\bf 131} (1963) 1801--1821.

\bibitem{Barashenkov:1972id}
V.~S. Barashenkov, H.~W. Bertini, K.~Chen, G.~Friedlander, G.~D. Harp, A.~S.
  Iljinov, J.~M. Miller, and V.~D. Toneev, {\it {Medium energy intranuclear
  cascade calculations - a comparative study}},  {\em Nucl. Phys.} {\bf A187}
  (1972) 531--544.

\bibitem{Bertini:1970zs}
H.~W. Bertini, {\it {Intranuclear-cascade calculation of the secondary nucleon
  spectra from nucleon-nucleus interactions in the energy range 340 to 2900 mev
  and comparisons with experiment}},  {\em Phys. Rev.} {\bf 188} (1969)
  1711--1730.

\bibitem{heikkinen2003bertini}
A.~Heikkinen, N.~Stepanov, and J.~P. Wellisch, {\it Bertini intra-nuclear
  cascade implementation in geant4},  {\em arXiv preprint nucl-th/0306008}
  (2003).

\bibitem{wright2015geant4}
D.~Wright and M.~Kelsey, {\it The geant4 bertini cascade},  {\em Nuclear
  Instruments and Methods in Physics Research Section A: Accelerators,
  Spectrometers, Detectors and Associated Equipment} {\bf 804} (2015) 175--188.

\bibitem{KUDRYAVTSEV2009339}
V.~Kudryavtsev, {\it Muon simulation codes music and musun for underground
  physics},  {\em Computer Physics Communications} {\bf 180} (2009), no.~3 339
  -- 346.

\bibitem{Aharmim_2009}
B.~Aharmim, S.~N. Ahmed, T.~C. Andersen, A.~E. Anthony, N.~Barros, E.~W. Beier,
  A.~Bellerive, B.~Beltran, M.~Bergevin, S.~D. Biller, and et~al., {\it
  Measurement of the cosmic ray and neutrino-induced muon flux at the sudbury
  neutrino observatory},  {\em Physical Review D} {\bf 80} (Jul, 2009).

\bibitem{Agostini:2018fnx}
{\bf Borexino} Collaboration, M.~Agostini et~al., {\it {Modulations of the
  Cosmic Muon Signal in Ten Years of Borexino Data}},  {\em JCAP} {\bf 1902}
  (2019) 046, [\href{http://xxx.lanl.gov/abs/1808.04207}{{\tt
  arXiv:1808.04207}}].

\bibitem{Javi:thesis}
J.~Mu\~noz Vidal, {\em {The NEXT path to neutrino inverse hierarchy}}.
\newblock PhD thesis, Valencia U., IFIC, 2017.

\bibitem{wittenberg1986lunar}
L.~Wittenberg, J.~Santarius, and G.~Kulcinski, {\it Lunar source of 3he for
  commercial fusion power},  {\em Fusion technology} {\bf 10} (1986), no.~2
  167--178.

\bibitem{bilder2009legal}
R.~B. Bilder, {\it A legal regime for the mining of helium-3 on the moon: Us
  policy options},  {\em Fordham Int'l LJ} {\bf 33} (2009) 243.

\bibitem{taylor1994helium}
L.~A. Taylor, {\it Helium-3 on the moon: model assumptions and abundances},  in
  {\em Engineering, Construction, and Operations in Space IV}, pp.~678--686,
  ASCE, 1994.

\bibitem{isotope2016defining}
C.~A. Slocum, {\it Defining the helium-3 industry for private sector: Current
  and projected resources, supply/demand, processing and transportation of the
  critical mineral isotope 3-he}, .

\bibitem{Ahmad:2002jz}
{\bf SNO} Collaboration, Q.~R. Ahmad et~al., {\it {Direct evidence for neutrino
  flavor transformation from neutral current interactions in the Sudbury
  Neutrino Observatory}},  {\em Phys. Rev. Lett.} {\bf 89} (2002) 011301,
  [\href{http://xxx.lanl.gov/abs/nucl-ex/0204008}{{\tt nucl-ex/0204008}}].

\bibitem{amsbaugh2007array}
J.~F. Amsbaugh, J.~Anaya, J.~Banar, T.~Bowles, M.~Browne, T.~Bullard,
  T.~Burritt, G.~Cox-Mobrand, X.~Dai, H.~Deng, et~al., {\it An array of
  low-background 3he proportional counters for the sudbury neutrino
  observatory},  {\em Nuclear Instruments and Methods in Physics Research
  Section A: Accelerators, Spectrometers, Detectors and Associated Equipment}
  {\bf 579} (2007), no.~3 1054--1080.

\bibitem{shea2010helium}
D.~Shea and D.~Morgan, {\it The helium-3 shortage: Supply, demand, and options
  for congress},  {\em Congressional Research Service} (01, 2011) R41419.

\bibitem{tastevin2000optically}
G.~Tastevin, {\it Optically polarized helium-3 for nmr imaging in medicine},
  {\em Physica scripta} {\bf 2000} (2000), no.~T86 46.

\bibitem{fain2010imaging}
S.~Fain, M.~L. Schiebler, D.~G. McCormack, and G.~Parraga, {\it Imaging of lung
  function using hyperpolarized helium-3 magnetic resonance imaging: review of
  current and emerging translational methods and applications},  {\em Journal
  of magnetic resonance imaging} {\bf 32} (2010), no.~6 1398--1408.

\bibitem{korff1939neutron}
S.~Korff and W.~Danforth, {\it Neutron measurements with boron-trifluoride
  counters},  {\em Physical Review} {\bf 55} (1939), no.~10 980.

\bibitem{lintereur20113he}
A.~Lintereur, K.~Conlin, J.~Ely, L.~Erikson, R.~Kouzes, E.~Siciliano,
  D.~Stromswold, and M.~Woodring, {\it 3he and bf3 neutron detector pressure
  effect and model comparison},  {\em Nuclear Instruments and Methods in
  Physics Research Section A: Accelerators, Spectrometers, Detectors and
  Associated Equipment} {\bf 652} (2011), no.~1 347--350.

\bibitem{goodings1972neutron}
A.~Goodings and J.~W. Leake, {\it Neutron counter filled with boron trifluoride
  gas},  Nov.~7, 1972.
\newblock US Patent 3,702,409.

\bibitem{fowler1950boron}
I.~Fowler and P.~Tunnicliffe, {\it Boron trifluoride proportional counters},
  {\em Review of Scientific Instruments} {\bf 21} (1950), no.~8 734--740.

\bibitem{segre1947boron}
E.~Segr{\`e} and C.~Wiegand, {\it Boron trifluoride neutron detector for low
  neutron intensities},  {\em Review of Scientific Instruments} {\bf 18}
  (1947), no.~2 86--89.

\bibitem{BOHLEN2014211}
T.~B{\"o}hlen, F.~Cerutti, M.~Chin, A.~Fass{\`o}, A.~Ferrari, P.~Ortega,
  A.~Mairani, P.~Sala, G.~Smirnov, and V.~Vlachoudis, {\it The fluka code:
  Developments and challenges for high energy and medical applications},  {\em
  Nuclear Data Sheets} {\bf 120} (2014) 211 -- 214.

\bibitem{Ferrari:2005zk}
A.~Ferrari, P.~R. Sala, A.~Fass{\`o}, and J.~Ranft, {\it {FLUKA: A
  multi-particle transport code (Program version 2005)}}, .

\end{thebibliography}\endgroup


\providecommand{\href}[2]{#2}\begingroup\raggedright\endgroup
